\PassOptionsToPackage{svgnames}{xcolor}
\documentclass[aip,jcp,amsmath,amssymb,reprint]{revtex4-2}
\usepackage{graphicx}
\usepackage{dcolumn}
\usepackage{bm}
\usepackage[usenames,dvipsnames,svgnames]{xcolor}
\usepackage{pstricks}
\usepackage{comment}
\usepackage{hyperref}

\definecolor{BLACK}{RGB}{0,0,0}
\newcommand{\Revision}[1]{\textcolor{black}{{#1}}}

\begin{document}

\title{\Revision{Breakdown of the Stokes-Einstein relation in Stillinger-Weber Silicon}}
\author{Himani Rautela}
\author{Shiladitya Sengupta$^*$}
\affiliation{Dept. of Physics, Indian Institute of Technology Roorkee, 247667 Uttarakhand, India.}
\email{shiladityasg@ph.iitr.ac.in}
\author{Vishwas V. Vasisht$^*$}
\affiliation{Dept. of Physics,  Indian Institute of Technology Palakkad, Nila Campus, Kanjikode, Palakkad, Kerala, India 678623.}
\email{vishwas@iitpkd.ac.in}

\date{\today}


\begin{abstract}
{\color{black}
	We investigate the dynamical properties of liquid and supercooled liquid silicon, modeled using the Stillinger-Weber (SW) potential, to examine the validity of the Stokes-Einstein (SE) relation. Towards this end, we examine the relationship among various dynamical quantities, including (i) the macroscopic transport coefficients - self diffusion coefficient $D$ and viscosity $\eta$, (ii) relaxation time $\tau_{\alpha}$ as well as (iii) lengthscale dependent relaxation times $\tau_{\alpha}(q)$ over a broad range of temperature $T$, pressure $P$ and density $\rho$ covering both equilibrium and metastable liquid state points in the phase diagram. Our study shows a weak break down in SE relation involving $D$ and $\eta$, and the loci of \Revision{the breakdown of the SE relation} (SEB) is found in the high T liquid phase. The $\tau_{\alpha}$, when used as a proxy to $\eta$, shows distinct breakdown in the SE relation whose loci is found in the supercooled liquid phase. Interestingly, certain parts of the phase diagram shows that loci of onset of slow dynamics lie below the loci of SEB, suggesting a regime that exhibits Arrhenius but non-Fickian behaviour. Computation of $\tau_{\alpha}(q)$ enables us to extract the lengthscale associated with the Fickian to non-Fickian behaviour using which we show that the \Revision{breakdown of the SE relation} occurs only below a specific lengthscale at a given temperature. Further we also compare the SEB loci with other features of the phase behaviour, including the loci of compressiblity maximum, density maximum as well as diffusivity maximum.}
\end{abstract}

\maketitle

\section{Introduction} \label{sec:intro}

{\color{black} Equilibrium liquid, when cooled faster than its typical nucleation time, remains in the liquid state even below its freezing point. This state, which is metastable relative to the crystalline phase, is referred to as a supercooled liquid. Such liquids exhibit rich physics in terms structural, dynamical and thermodynamic properties \cite{DebenedettiBook}. In particular, the dynamics of supercooled liquids show certain hallmark features such as the "decoupling of time scales" - long-time structural relaxation ($\alpha$-process) being characterized by \emph{multiple} independent timescales, which in turn has implications on fundamental aspect of fluctuation-dissipation theorem represented in terms of the Stokes-Einstein relation (SER). The SER was originally derived for a Brownian particle of radius $R$, at a given temperature $T$, diffusing in a homogeneous fluid having a drag coefficient $\zeta$. The diffusion coefficient of the Brownian particle follow the relation $D = \frac{k_B T}{\zeta}$, termed as the Sutherland-Einstein-Smoluchowski equation. Using the Stokes's law for the drag coefficient $\zeta = c \pi R \eta$, with c being a constant \cite{2019Costigliola} and $\eta$ being the viscosity of the background fluid, one obtains the celebrated Stokes-Einstein relation \cite{Sutherland1905, Einstein1905, Smoluchowski1906, LandauFMBook} given by:
\begin{align}
D = {k_B T \over c \pi R \eta} \propto ({\eta \over T})^{-1}.
\label{eqn:SER}
\end{align}
Both in experiments as well as in simulation studies \cite{1993Hodgdon}, it has been found that Eqn. \ref{eqn:SER} remains valid at high temperature equilibrium liquids even when $D$ is the \emph{self} diffusion coefficient, up to atomic length-scales \cite{HansenMcdonald}. Often, with the decrease in temperature, a breakdown in this relation has been observed and ratio $D \eta / T$ found to be temperature dependent instead of being a constant \cite{1992Fujara, 1994Chang}. The extent of breakdown can be also expressed in term a fractional SER  defined as:}
\begin{align}
    D \propto ({\eta \over T})^{-\xi}.
    \label{eqn:fracSE}
\end{align}
\noindent where the fractional $\xi$ exponent is $1$ if the SER is valid or $0 < \xi <1$ in case of breakdown (SEB).

The SER is typically obeyed at high temperatures, where the Fick's law of diffusion is valid: the current $\vec{j}(\vec{r},t)$ of local single particle density $\rho(\vec{r},t)$ is linearly proportional to the gradient of $\rho$, $\vec{j} = -D \nabla \rho$ \cite{HansenMcdonald}. This regime, termed as  "Fickian" regime, is characterized by a single time scale of Fickian diffusion. A high temperature liquid is also structurally homogeneous beyond single particle lengthscale  \cite{HansenMcdonald}.  However with lowering of temperature, as the liquid enters the metastable equilibrium, with growing many-body correlation the dynamics becomes spatially heterogeneous \cite{2005Sciortino, 2006Dyre, Banerjee2017, 2020Tanaka}. This phenomenon is known as ``dynamical heterogeneity'' (DH) \cite{DHBook}. Concurrently, various timescales of supercooled liquid dynamics get separated into two families - one proportional to the cooperative motion of the mobile particles, and the second depicting the dynamics of the immobile particles \cite{Kawasaki2019, Das2022}. Thus, the supercooled liquid regime shows ``non-Fickian'' dynamics and the SEB \cite{TK1995, CE1996, Silescu1999, Ediger2000, 2015Mishra}. As noted above, DH provides an intuitive physical mechanism to understand the origin of the non-Fickian nature of the dynamics \cite{Ediger2000}. However, we note that it is not the only possibility. For example, some liquids show ``Fickian but non-Gaussian'' dynamics \cite{2017Acharya} which may be explained by a stochastically varying ``diffusing diffusivity'' model \cite{2016Jain}.

\begin{figure}[htbp]
  \centering
  \includegraphics[width=0.45\textwidth]{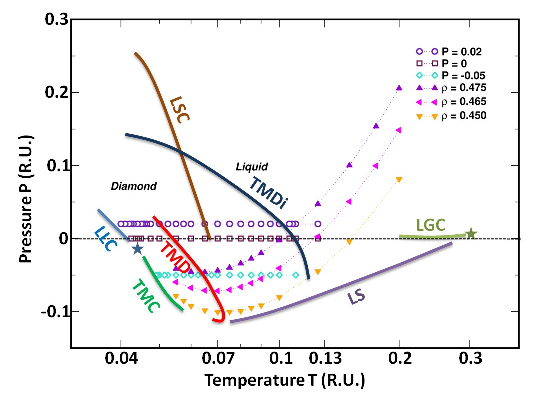}
  \caption{\emph{Phase diagram in the (P,T) plane of the Stillinger-Weber silicon model}. Boundaries of the stable liquid phase is shown by (i) the liquid-solid (LSC) (ii) the liquid-gas coexistence (LGC) lines along with liquid-gas critical point (light green star), and (iii) the liquid spinodal (LS) line.  Also shown are the liquid-liquid coexistence (LLC) line and associated liquid-liquid critical point (blue star) deep in the supercooled regime. The region of anomaly is marked by the loci of diffusivity (TMDi), density (TMD) and compressibility maxima (TMC or the Widom line). Dotted lines along with the state points are the isobars and isochores analyzed in the present study. \Revision{The loci are schematic adaptations from Refs. N. Phys. 7, 549 (2011) \cite{Vasisht2011}, Copyright 2011, Springer Nature Limited   and ``Liquid Polymorphism: Advances in Chemical Physics'', Vol. 152 (2013), Ed. H. Eugene Stanley \cite{Vasisht2013}, Copyright 2013 John Wiley \& Sons, Inc.} }
    \label{fig:phaseDiag_SWSi}
\end{figure}

Compared to simple liquids, network-forming liquids like silicon, silica and water exhibit a richer phase behaviour. Here the inter-particle interaction favors a 4-coordinated, tetrahedrally ordered, low density local structure as opposed to a 5-coordinated, disordered local environment with higher density. Hence, on one hand, with decreasing temperature, particles tend to form an energetically stable tetrahedral network, and on the other hand, with increasing pressure (or density), the more densely packed disordered configuration is preferred \cite{2017Handle}. This competition has two important consequences: (i) network-forming liquids can display a novel liquid-liquid phase transition and associated liquid-liquid critical point \cite{Poole1992, Vasisht2011, Vasisht2013} and (ii) they can exhibit anomalies in various structural, dynamic and thermodynamic properties \cite{Errington2001, Errington2006, Shell2002, Hujo2011, Vasisht2014, Sengupta2014}. Fig. \ref{fig:phaseDiag_SWSi} illustrates the computed $P-T$ phase diagram for silicon \cite{Vasisht2011, Vasisht2014}. It shows (i) ``LSC'', the liquid-solid and (ii) ``LGC'' the liquid-gas co-existence lines and the liquid-gas critical point; (iii) ``LLC'', the liquid-liquid co-existence line ending in a critical point, existing deep inside the supercooled liquid regime; (iv) ``LS'', the liquid spinodal beyond which the liquid is mechanically unstable and (v) loci of maxima of dynamic, thermodynamic and structural anomalies described respectively by (a) ``TMDi'', temperature of diffusivity maxima, (b) ``TMD'', temperature of density maxima, and (c) ``TMC'', temperature of compressibility maxima (Widom line). The loci of TMDi, TMD and TMC marks the anomalous region of the phase diagram. We note that the different anomalies are considered to be related to one another to form a nested structure \cite{Errington2001, Vasisht2014}. Thus, in silicon, on top of the slowdown of dynamics with cooling, there is an additional effect due to the anisotropy of interaction leading to local structural heterogeneity and hence anomalies.

In this study we focus on validity of the SER in model silicon using atomistic simulation. It is motivated by two distinct lines of research. First, understanding the effect of the anomalies \cite{2000Angell, 2020Shi, Gallo2021} on the nature of dynamics in network-forming liquids. In water the SEB were observed to be in the vicinity of the Widom line \cite{Kumar2007}, suggesting significant role of anomalies. However, there are also indications that in network-forming \cite{Micoulaut2016} and other class \cite{Wei2018} of complex liquids the SEB may occur in the normal liquid regime itself. In liquids with isotropic interactions, {\it e.g.}, the three-dimensional Kob-Andersen model, it has been shown that $T_{SEB} \approx T_{onset} \approx T_m$ \cite{Sengupta2013, Banerjee2017, 2018Pedersen}, where $T_{m}$ is the melting temperature, {\it i.e.} the SER is valid in the normal liquid, and the SEB occurs in the supercooled liquid. Hence it is useful to identify the loci of $T_{SEB}$ on the phase diagram. This task is however, made difficult by the fact that often in literature, usually for computational efficiency, one dynamical measure is substituted for another. {\it E.g.} both $\tau_{\alpha}$ \cite{2006Chen} and $\tau_{\alpha}/T$ \cite{Xu2009} are used as proxies for $\eta$. However, such substitutions are not universally agreed upon, {\it e.g.} coupling of both $\eta$ \cite{Ediger2000} and $\frac{\eta}{T}$ \cite{2007Alonso} with $D$ are studied to verify the SER. Thus a systematic analysis of the relationship among different dynamical measures is necessary \cite{Shi2013, KK2017, 2023Pareek}.
Our second motivation is understanding the origin of the SEB in silicon. In simple liquids, the dominant point of view is that DH with a growing lengthscale leads to a lengthscale dependent cross-over from Fickian to non-Fickian dynamics and the SEB \cite{DHBook}. If the same is true in network-forming liquids, it will be interesting to extract the associated lengthscale. Note that in this picture, the SEB may occur even without any influence from anomalies.

To achieve above objectives, (i) we compute dynamical properties of liquid silicon modelled by the Stillinger-Weber (SW) potential \cite{SW1985} over a wide range of state points, characterize the SER and chart out the loci of its breakdown on the phase diagram. We also provide a systematic numerical comparison of the relationship among different timescales. (ii) We probe dynamics at different wavenumbers ($q$) to identify Fickian to non-Fickian crossover in dynamics and the lengthscale associated with it, to understand the origin of the observed SEB.

We report the following main results for silicon in this work. (a) We identify a regime in the $P-T$ phase diagram that is Arrhenius but non-Fickian. (b) Analyzing the $q$ dependence of the high temperature activation energy, we demonstrate that in silicon, the Arrhenius regime is not dynamically homogeneous (Fickian) on a timescale of $\tau_\alpha$. (c) We extract a growing dynamical lengthscale that characterizes the Fickian to non-Fickian crossover in dynamics. Thus we show the significant influence of DH in the SEB in silicon. 

The paper is organized as follows: in sec. \ref{sec:method} we give computational details and definitions. We present a comprehensive study of the SER and SEB in SW silicon in sec. \ref{sec:SEB}, and analysis of the Fickian to non-Fickian crossover in dynamics in sec \ref{sec:qdep}. Finally we summarize our findings and discuss the conclusions in sec. \ref{sec:conclusion}.


\section{Method} \label{sec:method}

\begin{figure*}[htbp]
    \centering
    \includegraphics[width=0.3\textwidth]{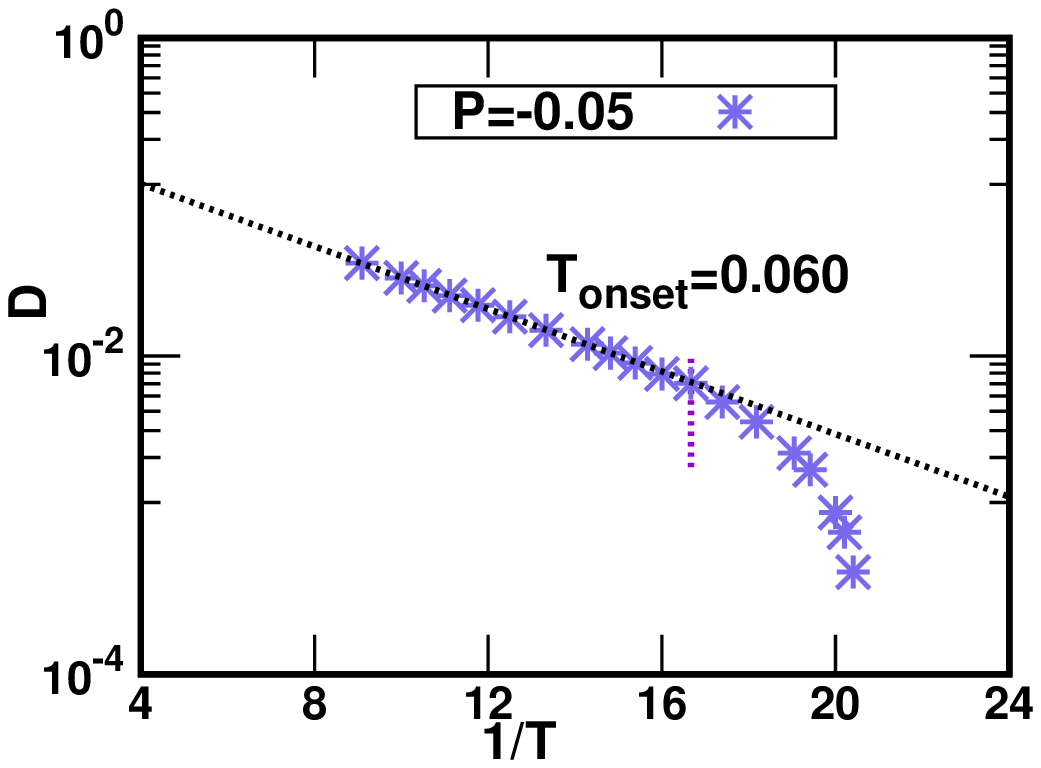}
    \includegraphics[width=0.3\textwidth]{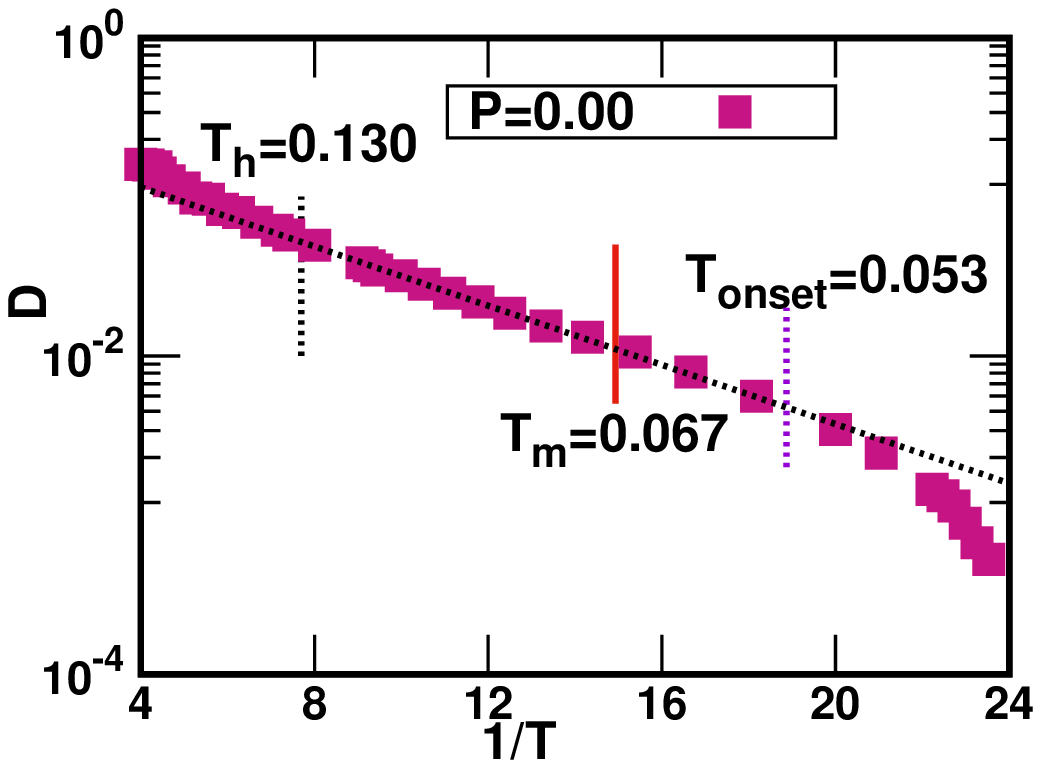}
    \includegraphics[width=0.3\textwidth]{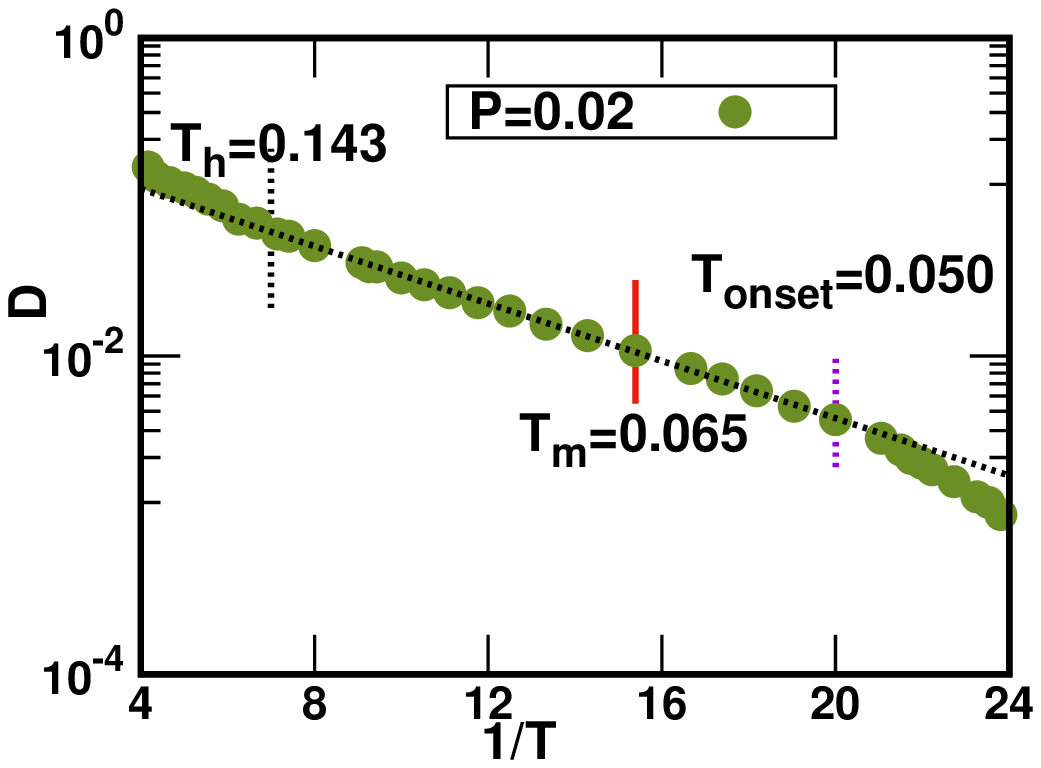}
    \put(-440,40){\textbf{(a)}}
    \put(-280,40){\textbf{(b)}}
    \put(-120,40){\textbf{(c)}}
    \\
    \includegraphics[width=0.3\textwidth]{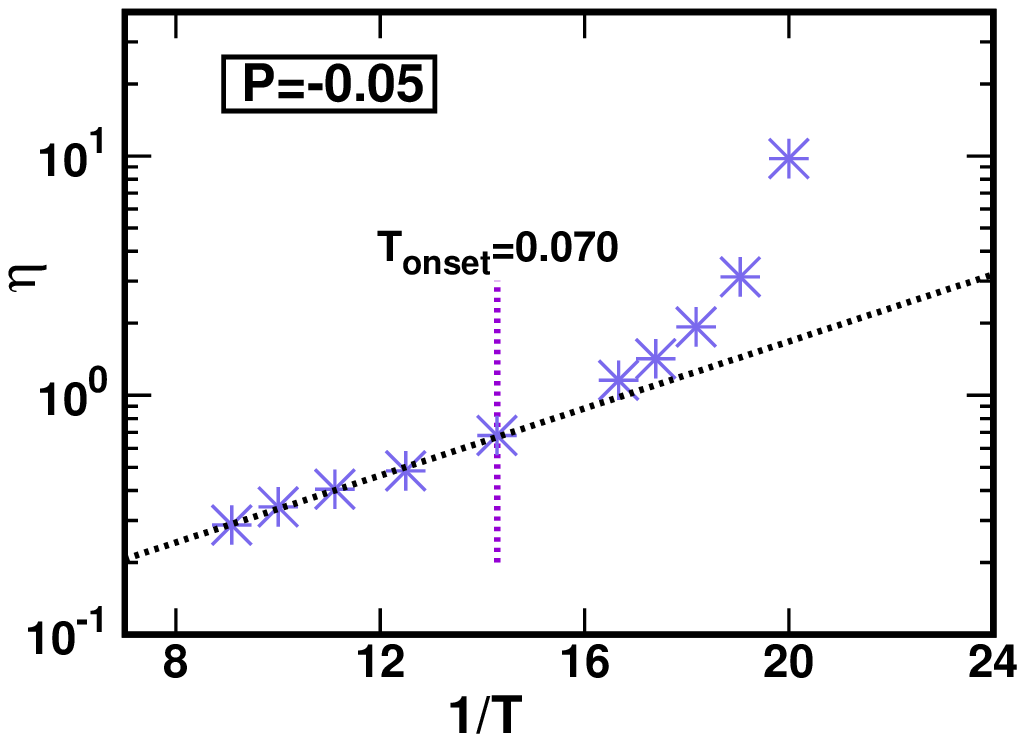}
    \includegraphics[width=0.3\textwidth]{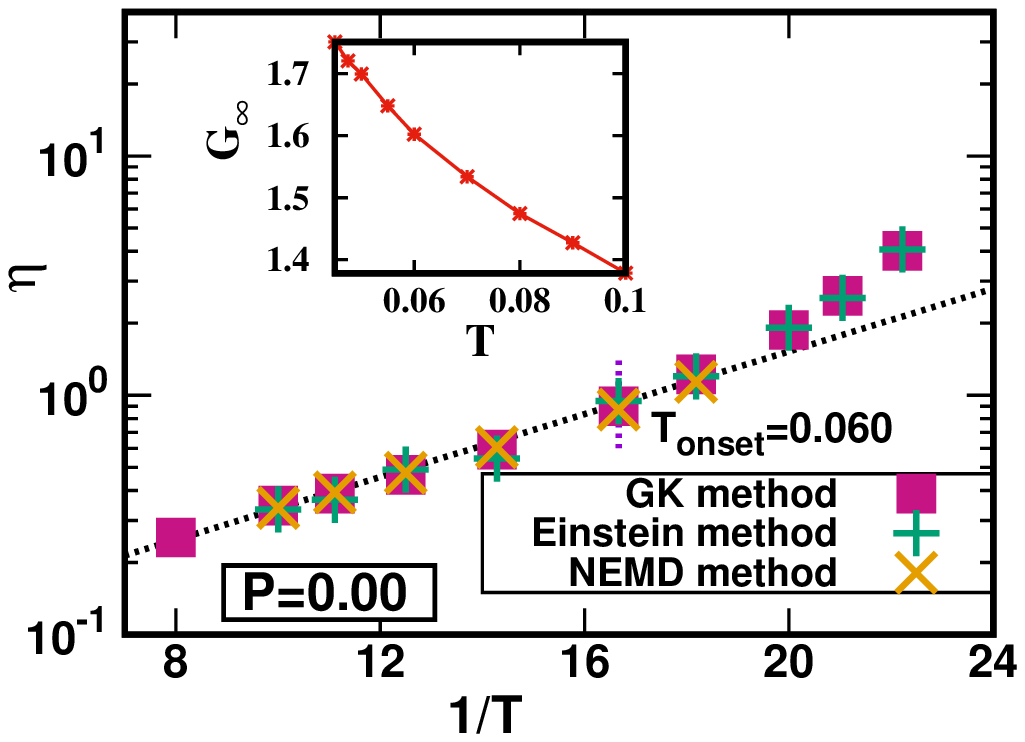}
    \includegraphics[width=0.3\textwidth]{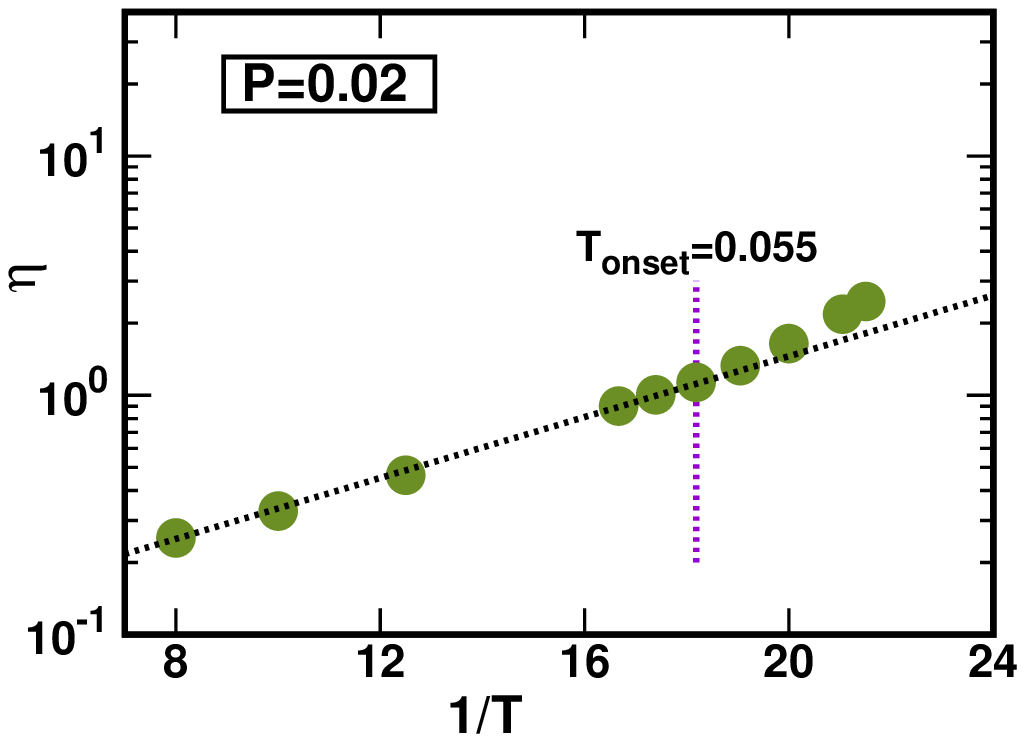}
    \put(-440,50){\textbf{(d)}}
    \put(-280,50){\textbf{(e)}}
    \put(-120,50){\textbf{(f)}}
    \\
    \includegraphics[width=0.3\textwidth]{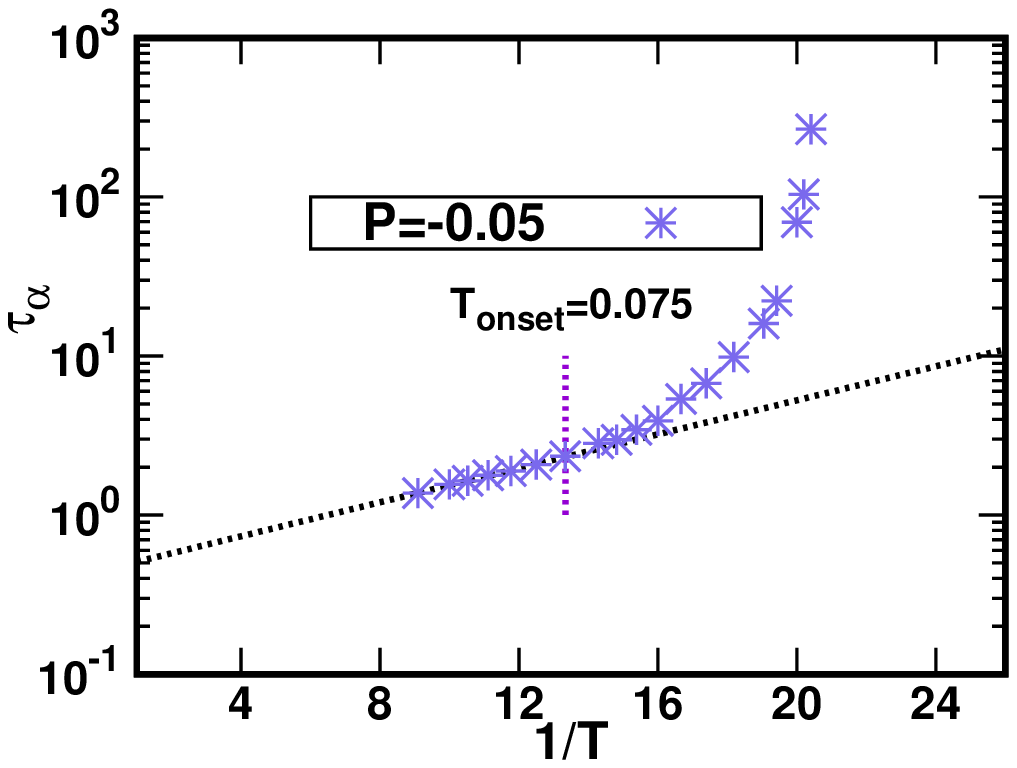}
    \includegraphics[width=0.3\textwidth]{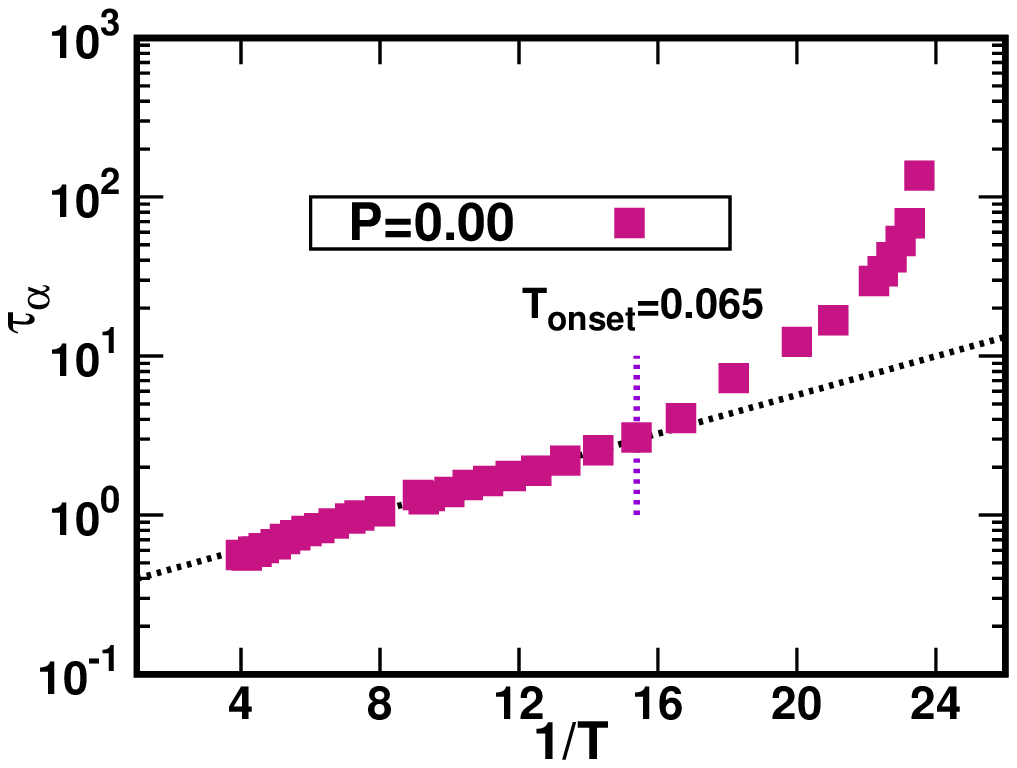}
    \includegraphics[width=0.3\textwidth]{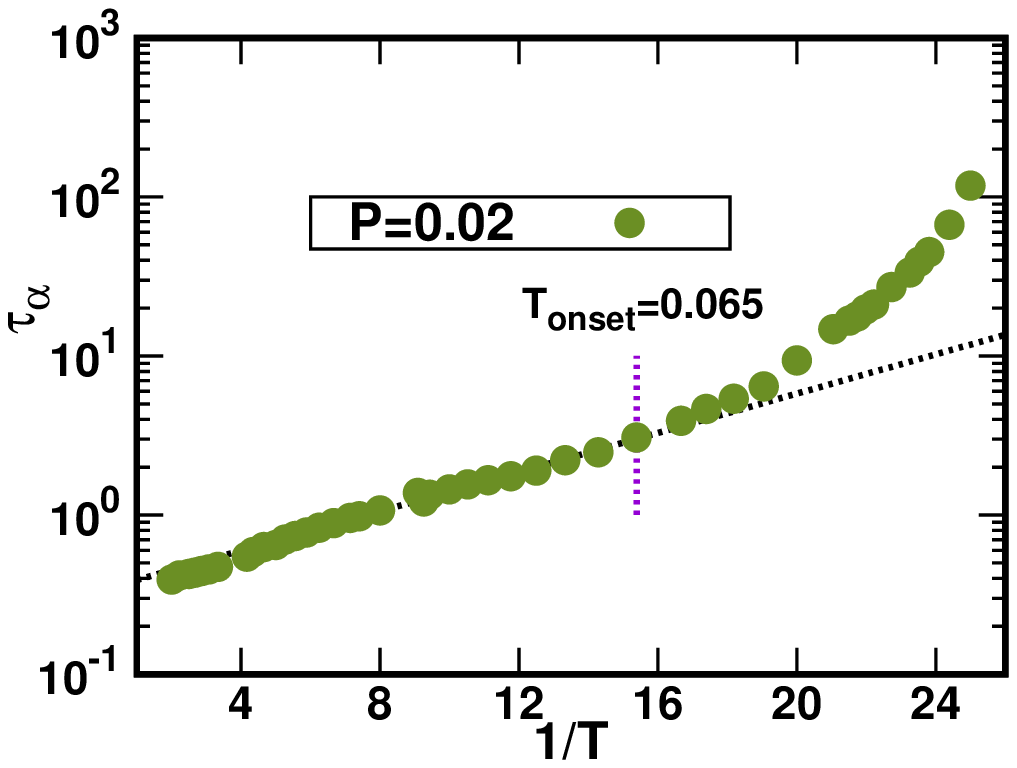}
    \put(-440,50){\textbf{(g)}}
    \put(-280,50){\textbf{(h)}}
    \put(-120,50){\textbf{(i)}}
    \caption{\emph{Timescales and characteristic temperatures along isobars of SW silicon.} We show the temperature ($T$) dependence of \textbf{(a)-(c)} diffusion coefficient $D$, \textbf{(d)-(f)} shear viscosity $\eta$ and \textbf{(g)-(i)} $\alpha$-relaxation time $\tau_\alpha$ along representative negative, zero and positive pressure isobars spanning a broad range of pressures on the phase diagram. Black dotted lines are fits to Arrhenius law to determine the onset temperature $T_{onset}$ below which a crossover to non-Arrhenius behavior occur. For diffusion coefficient, we also observe deviation from Arrhenius behavior above a temperature $T_h$. We limit ourselves to temperatures below $T_h$ for isobars. The melting temperatures $T_m$, available only for non-negative pressures, are also marked. \emph{Validation of viscosity calculation.} At $P=0$ isobar, panel \textbf{(e)}, the shear viscosity has been calculated using (i) Green-Kubo relation, Eqn \ref{eqn:GK} (GK method), (ii) Einstein relation, Eqn \ref{eqn:Einstein} (Einstein method) and (iii) Non-equilibrium molecular dynamics, Eqn \ref{eqn:MP} (NEMD method). Excellent agreement is obtained for all methods validating our calculation. \Revision{see also Fig. \ref{fig:eta_det} in the Appendix.} For other isobars and isochores, we present viscosity data only using the GK method. (\emph{Inset of \textbf{(e)}}): Temperature dependence of the instantaneous shear modulus $G_{\infty}$ showing a weak variation with $T$ and values $\sim \mathcal{O}{(1)}$. See Sec. \ref{sec:SEBtau} for more details.}
    \label{fig:TdepIsobar}
\end{figure*}

\begin{figure*}[htbp]
    \centering
     \includegraphics[width=0.3\textwidth]{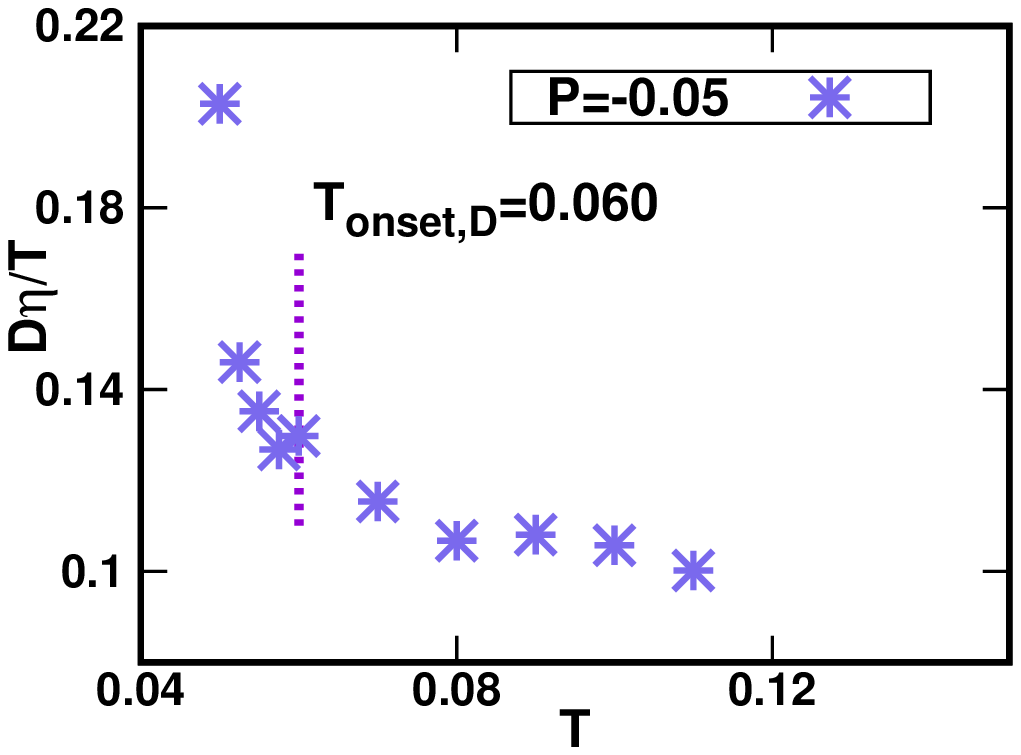}
     \includegraphics[width=0.3\textwidth]{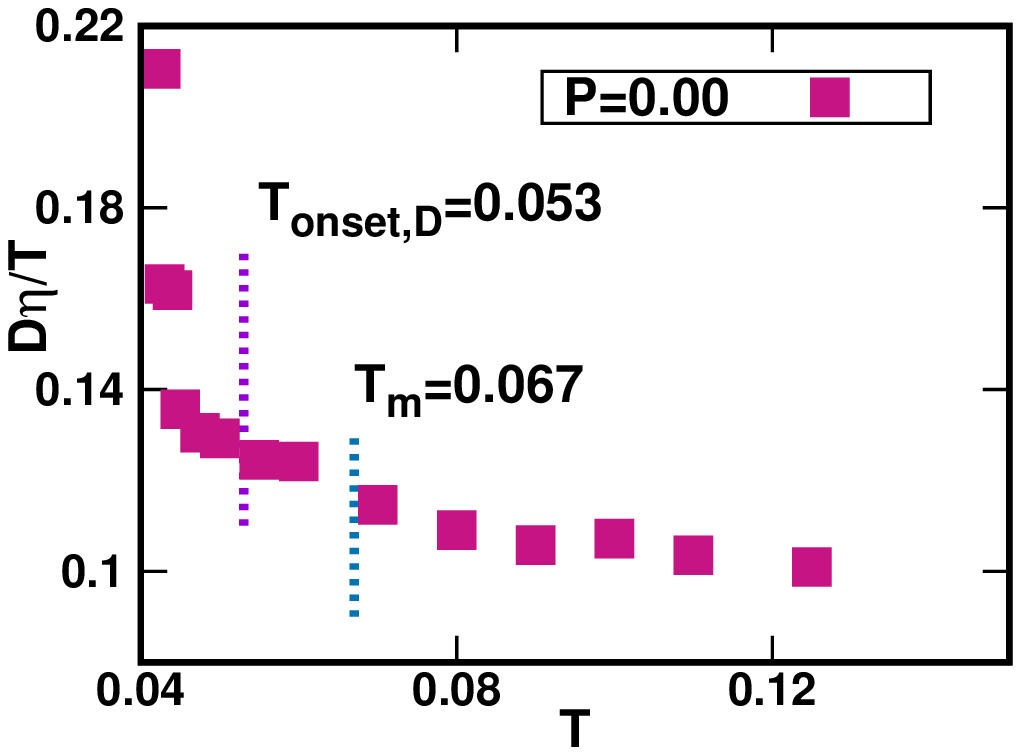}
     \includegraphics[width=0.3\textwidth]{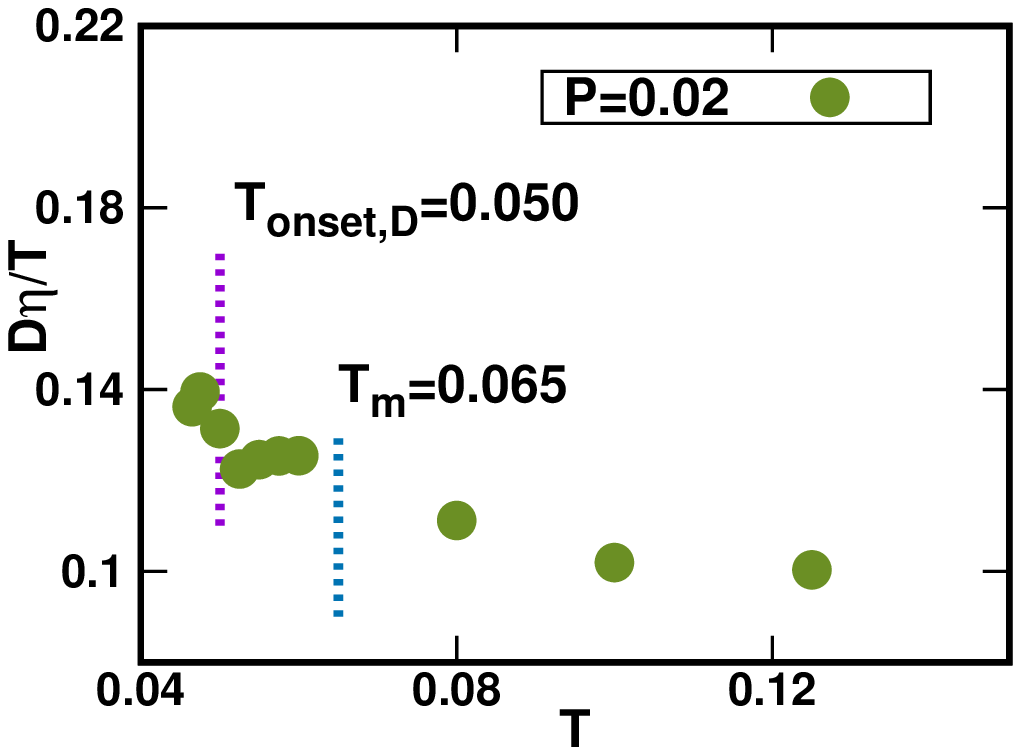}
     \put(-360,70){\textbf{(a)}}
     \put(-200,70){\textbf{(b)}}
     \put(-50,70){\textbf{(c)}}
     \\
     \includegraphics[width=0.3\textwidth]{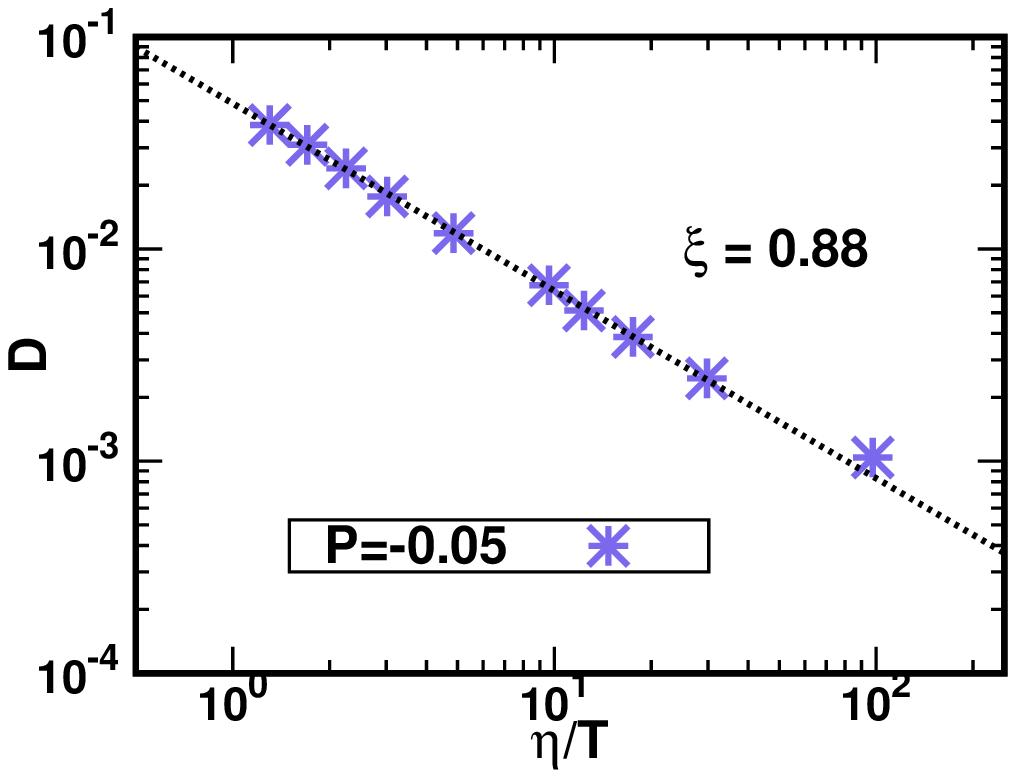}
     \includegraphics[width=0.3\textwidth]{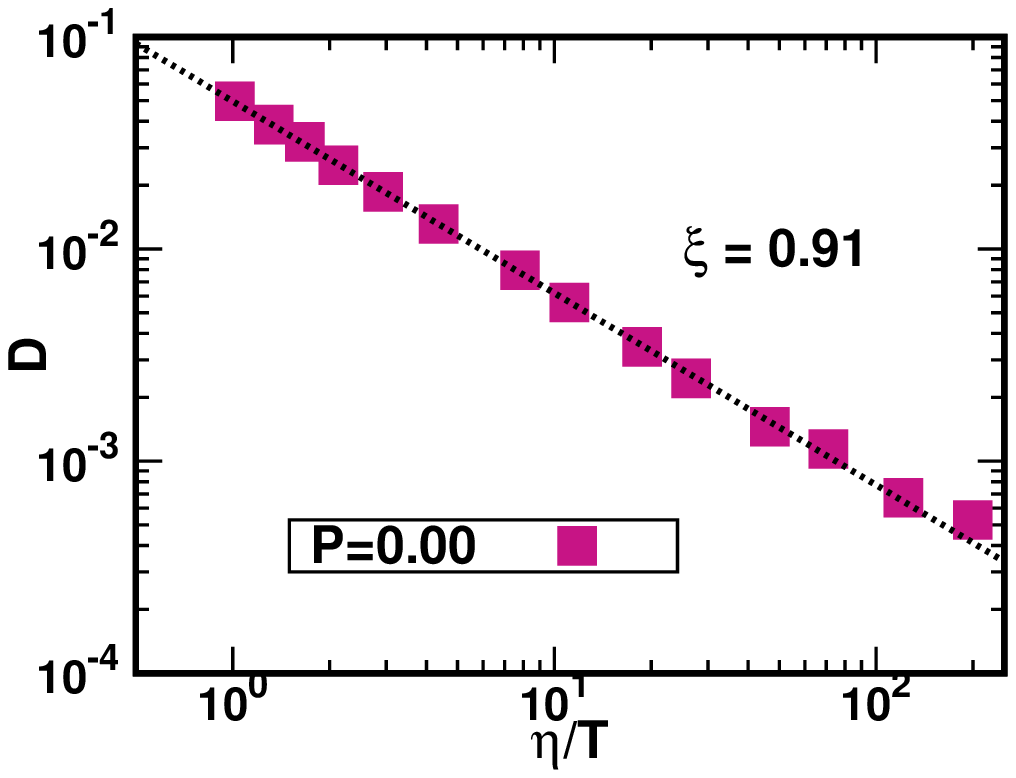}
     \includegraphics[width=0.3\textwidth]{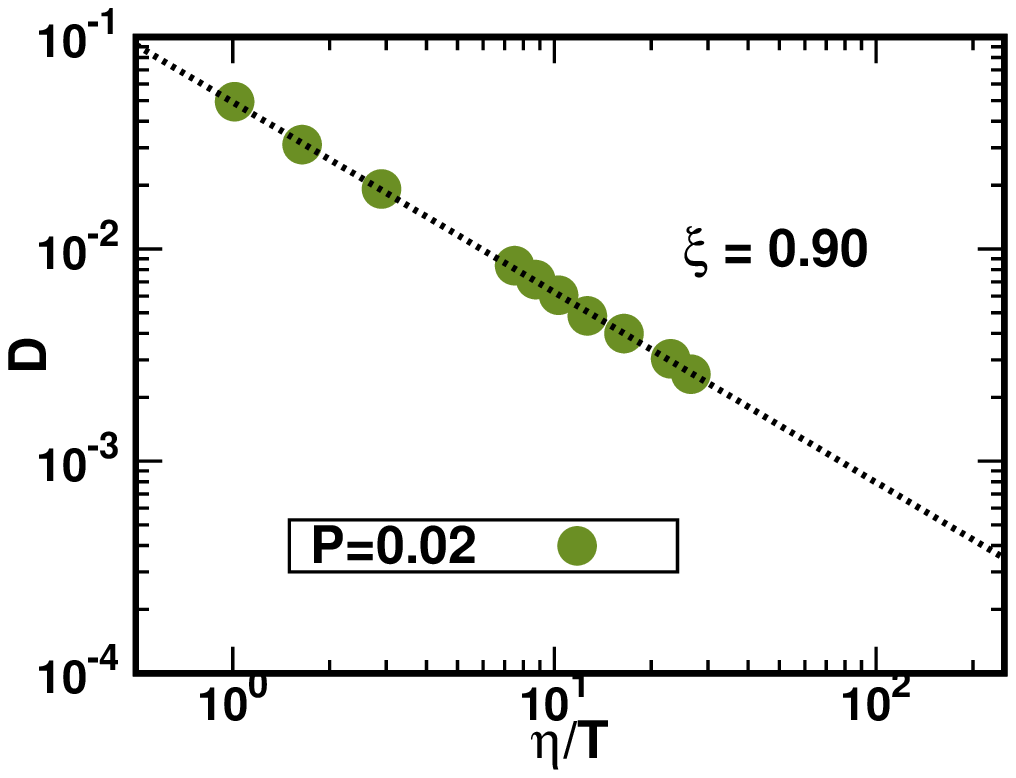}
     \put(-440,60){\textbf{(d)}}
     \put(-280,60){\textbf{(e)}}
     \put(-120,60){\textbf{(f)}}
     \caption{\emph{Testing the Stokes-Einstein relation along isobars of SW Silicon.} \textbf{(a)-(c)} The temperature variation of the SE ratio $\frac{D\eta}{T}$, along negative, zero and positive pressure isobars. It does not reach a constant value even above the melting temperature (within the Arrhenius regime {\it i.e.} upto $T_h$). \textbf{(d)-(f)} Plots of $D$ {\it vs.} $\frac{\eta}{T}$, show a fractional SE relation, Eqn. \ref{eqn:fracSE}, at all isobars studied. The fractional exponent $\xi$ values are $\approx 0.9$, see Table \ref{tab:isobar}. Thus SW Silicon exhibits a weak \Revision{SEB} along isobars.}
     \label{fig:SEBisobar}
\end{figure*}

\subsection{SW Silicon model}
The Stillinger-Weber potential \cite{SW1985} consists of a spherically symmetric two-body term $v_2$ and a direction dependent three-body term $v_3$, given by,
\begin{align}
  U_{SW} &= \sum_{i<j} v_2\Big(\frac{r_{ij}}{\sigma}\Big) + \lambda \sum_{i<j<k} v_3 \Big(\frac{\vec{r}_i}{\sigma},\frac{\vec{r}_j}{\sigma},\frac{\vec{r}_k}{\sigma}\Big) \nonumber\\
  v_2 (r) &= \begin{cases}
    A \epsilon (\frac{B}{r^4} - 1) \exp(\frac{1}{r-a}) & r < a,\\
    0 & r \geq a
  \end{cases} \nonumber\\
  v_3(\vec{r}_i,\vec{r}_j,\vec{r}_k) &= h(r_{ij},r_{ik},\theta_{jik}) + h(r_{ij},r_{jk},\theta_{ijk})  \nonumber\\
         & \quad + h(r_{ik},r_{jk},\theta_{ikj}) \nonumber\\
  h(r_{ij},r_{ik},\theta_{jik}) &= \epsilon \exp(\frac{\gamma}{r_{ij}-a}+\frac{\gamma}{r_{ik}-a})(\cos \theta_{jik}+\alpha)^2\nonumber\\
  &\times H(a-r_{ij})H(a-r_{ik})
\end{align}
where $\sigma$ is the diameter of an atom, $\vec{r}_i$ represents the position of $i$-th particle, $r_{ij}$ is the inter-particle separation between particles $i,j$, $\theta_{jik}$ is the angle between $\vec{r}_{ij}, \vec{r}_{ik}$ and $\lambda$ denotes the strength of the three-body term. The model was originally proposed for silicon, with $\lambda=21$ and $\alpha=\frac{1}{3}$ favoring tetrahedral bonding. The units of length and energy are $\sigma=2.0951 \mathring{A}$, $\epsilon = 209.5 \,kJ/mol$ respectively. Other parameters are same as in Refs. \cite{Vasisht2011, Sengupta2014}. We used both in-house code and LAMMPS package \cite{LAMMPS} for the molecular dynamics (MD) simulation of $N=512$ atoms using both isochoric-isothermal (NVT) and isobaric-isothermal (NPT) ensembles, with time step of $0.005$ in reduced units (r.u.). We report data for three isobars with $P=0.02, 0, -0.05$ (r.u.) and three isochores $\rho=0.450, 0.465, 0.475$ (r.u.) spanning a wide range of positive to negative pressure regime in the phase diagram, as shown in Fig. \ref{fig:phaseDiag_SWSi}. For each state point, the runlength of MD trajectories were $10-100\,\tau_{\alpha}$.

\subsection{Definitions}
\paragraph{Diffusion coefficient $D$:}  mean squared displacement (MSD) {\it vs.} time $t$ and the diffusion coefficient $D$ were computed using the following definitions, 
\begin{align}
    MSD (t) &= \left\langle \frac{1}{N}\sum_{i=1}^N\left[\vec{r}_i (t+t_0) - \vec{r}_i (t_0) \right]^2 \right\rangle\nonumber\\
    D &=\lim_{t\rightarrow\infty}\frac{MSD(t)}{6t}.
    \label{eqn:D}
\end{align} 
Here $i$ is particle index, $t_0$ denotes time origin, $\langle \cdot \rangle \equiv \frac{1}{N_{t_{0}}} \sum_{t_0=1}^{N_{t_{0}}} (\cdot)$ indicates ensemble averaging, and $N, N_{t_0}$ represent total number of particles, and time origins respectively.

\paragraph{Shear viscosity $\eta$:} 
In contrast to $D$ which is a single-particle quantity, $\eta$ is a many-body observable and hence is computationally much costlier. Consequently, the statistical uncertainty in $\eta$ is much higher. Hence for validation, we compute $\eta$ by three different methods:

(1) \emph{GK method}: From equilibrium stress-correlation function applying the Green-Kubo (GK) formula \cite{Haile},
\begin{align}\label{eqn:GK}
    P_{\alpha\beta}(t) &= \frac{1}{V}\left(\sum_{i=1}^N \frac{p_{i\alpha}p_{i\beta}}{m} + \sum_{i=1}^N\sum_{j>i}^N  r_{ij\alpha}f_{ij\beta}\right), \nonumber\\
    \langle P_{\alpha\beta}(t)P_{\alpha\beta}(0) \rangle &= \frac{1}{N_{t_{0}}} \sum_{t_0=1}^{N_{t_0}} P_{\alpha\beta}(t+t_0)P_{\alpha\beta}(t_0) \nonumber\\
    \eta &= \frac{V}{k_BT}\int_{0}^{\infty} dt\, \langle P_{\alpha\beta}(t)P_{\alpha\beta}(0) \rangle
\end{align}
where $\vec{r}_{i}(t), \vec{p}_i(t)$ are the position and momentum of particle $i$ at a time instant $t$, $\vec{r}_{ij}$ is the inter-particle separation, $\vec{f}_{ij}(t)$ is the force on particle $i$ by particle $j$, $\alpha, \beta$ are Cartesian components of coordinates, $P_{\alpha\beta} (t)$ is the stress tensor, $\langle P_{\alpha\beta}(t)P_{\alpha\beta}(0) \rangle$ is the stress auto-correlation function ensemble-averaged over $N_{t_0}$ time origins, $V$ is the volume. 

(2) \emph{Einstein method}: From the Helfand moment $A_{\alpha\beta}(t)$ of equilibrium stress tensor applying the Einstein relation \cite{Haile},
\begin{align}\label{eqn:Einstein}
    \frac{dA_{\alpha\beta}(t)}{dt} &= P_{\alpha\beta}(t) V \nonumber\\
    \langle [A_{\alpha\beta}(t) - A_{\alpha\beta}(0)]^2 \rangle &= \frac{1}{N_{t_{0}}} \sum_{t_0=1}^{N_{t_0}} [A_{\alpha\beta}(t+t_0) - A_{\alpha\beta}(t_0)]^2 \nonumber\\
    \eta &= \frac{1}{V k_BT} \lim_{t\rightarrow\infty}\frac{\langle [A_{\alpha\beta}(t) - A_{\alpha\beta}(0)]^2 \rangle}{2t}
\end{align}

(3) \emph{NEMD method}: From \emph{non}-equilibrium, reverse molecular dynamics simulation under an imposed momentum flux applying the defining constitutive relation for shear viscosity \cite{viscMP},
\begin{align}\label{eqn:MP}
    j_z (p_x) &= -\eta\frac{\partial v_x}{\partial z}
\end{align}
where $j_z$ represents a flux of transverse linear momentum due to an imposed velocity gradient field $\frac{\partial v_x}{\partial z}$. Note that as the method is based on linear response, the imposed velocity profile must be \emph{linear} in $z$.

In Fig. \ref{fig:TdepIsobar}(e) representative data at $P=0$ show that viscosity computed by all three methods has excellent agreement with each other, \Revision{see also Fig. \ref{fig:eta_det} in the Appendix.} For further analysis we report the viscosity computed by the GK method only.

\paragraph{$q$-dependent relaxation time $\tau(q)$:} The intermediate ($F(q,t)$) scattering functions were computed using \cite{KobBinder}, 
\begin{align}
    \rho_{\vec{q}}(t) &= \sum_{i=1}^N \exp(-\mathrm{i} \vec{q}\cdot \vec{r}_i(t))\nonumber\\
    F(q,t) &= \frac{1}{N} \langle \rho_{\vec{q}}(t)\rho_{-\vec{q}}(0)\rangle 
    \label{eqn:tauqFqt}
\end{align} 
The wave number $q$ dependent relaxation time $\tau(q)$ was calculated from the decay of $F(q,t)$ for a wide range of $q$ values using the definition $F(q,\tau) = \frac{1}{e}$. The $\alpha$-relaxation time $\tau_{\alpha}$ was computed at $q^*$ (first peak of static structure factor $S(q)$). We also consider longer timescales $\tau_{0.25} (q)$ and $\tau_{0.1} (q)$ defined as $F(q,\tau_{0.25})=0.25$ and $F(q,\tau_{0.1})=0.1$ respectively.


\section{\Revision{Loci of the Breakdown of the SE relation in SW Silicon}}\label{sec:SEB}

\subsubsection{Weak SEB along isobars}

\begin{figure*}[htbp]
    \centering
    \includegraphics[width=0.3\textwidth]{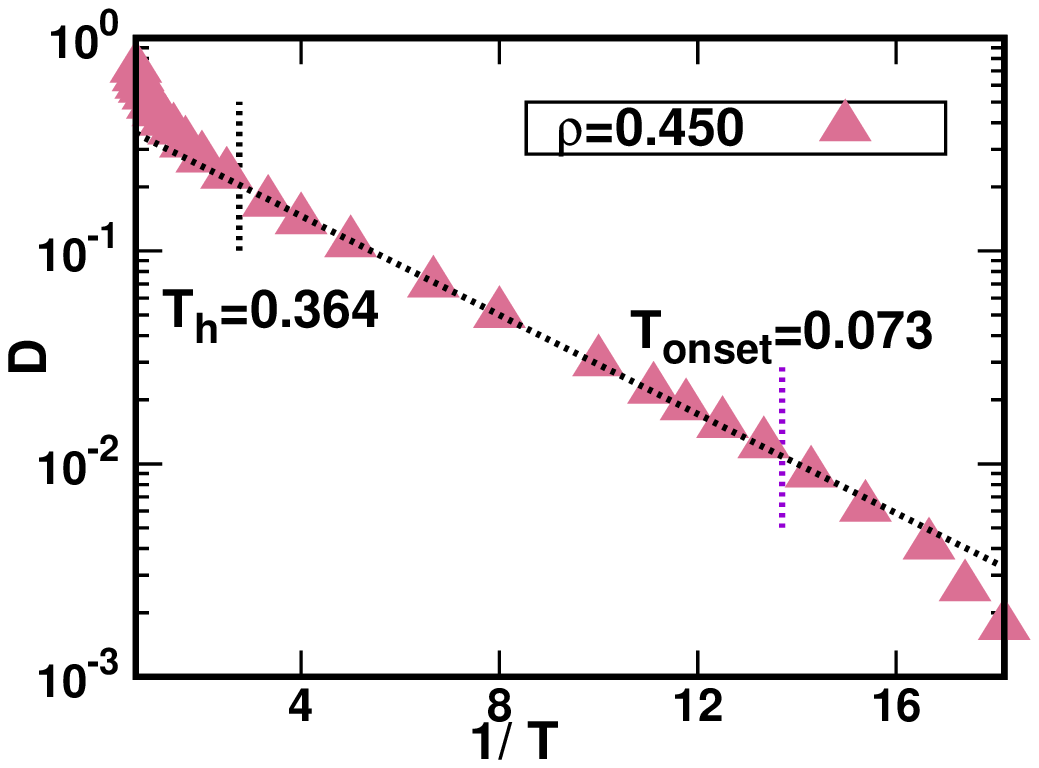}
    \includegraphics[width=0.3\textwidth]{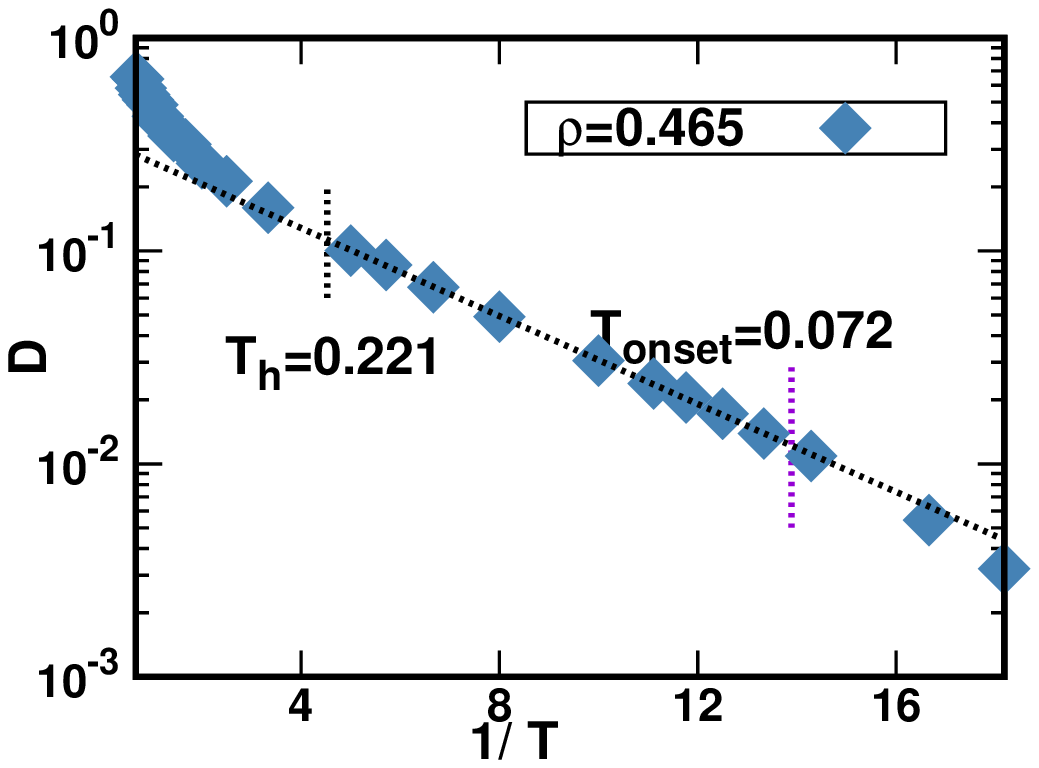}
    \includegraphics[width=0.3\textwidth]{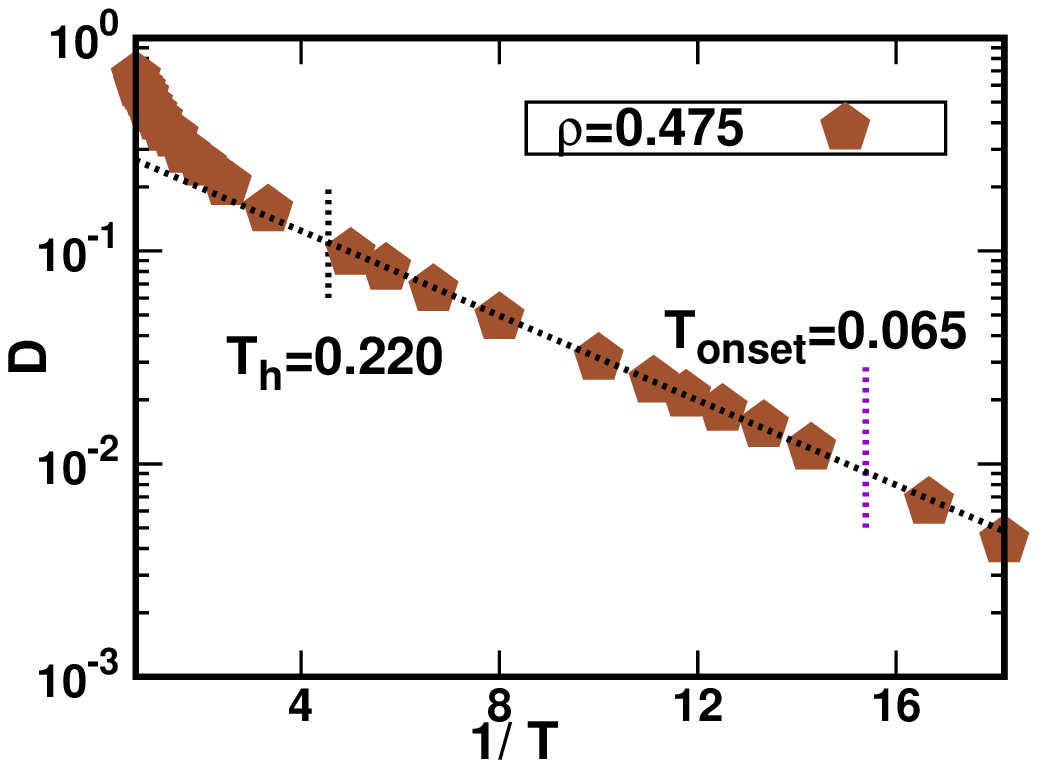}
    \put(-410,30){\textbf{(a)}}
    \put(-240,30){\textbf{(b)}}
    \put(-100,30){\textbf{(c)}}
    \\
    \includegraphics[width=0.3\textwidth]{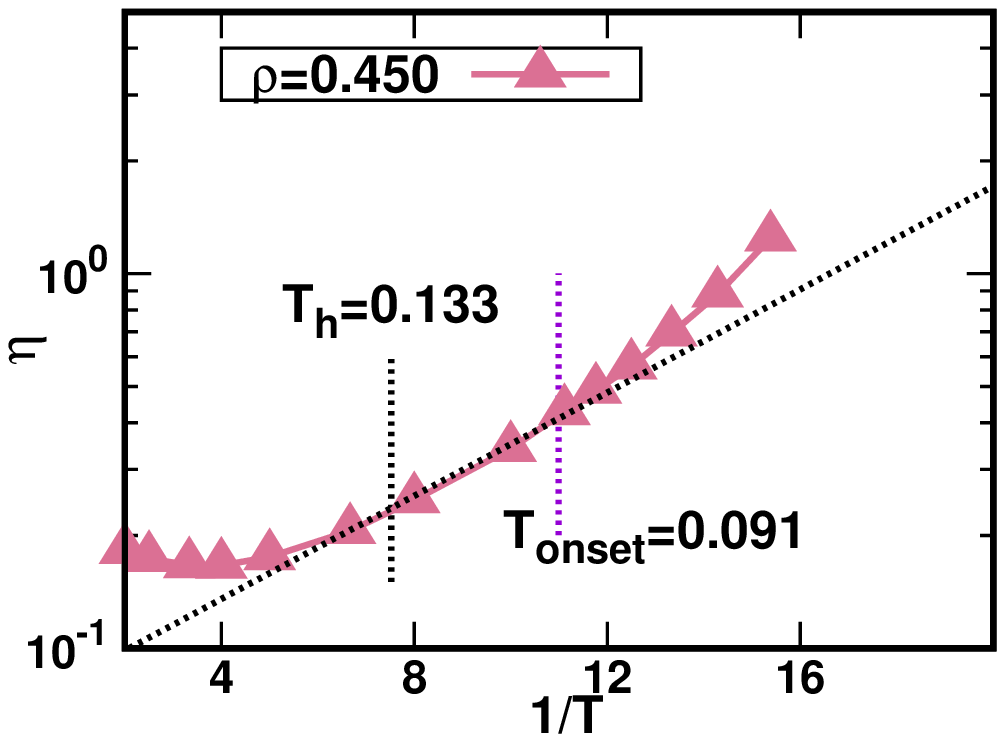}
    \includegraphics[width=0.3\textwidth]{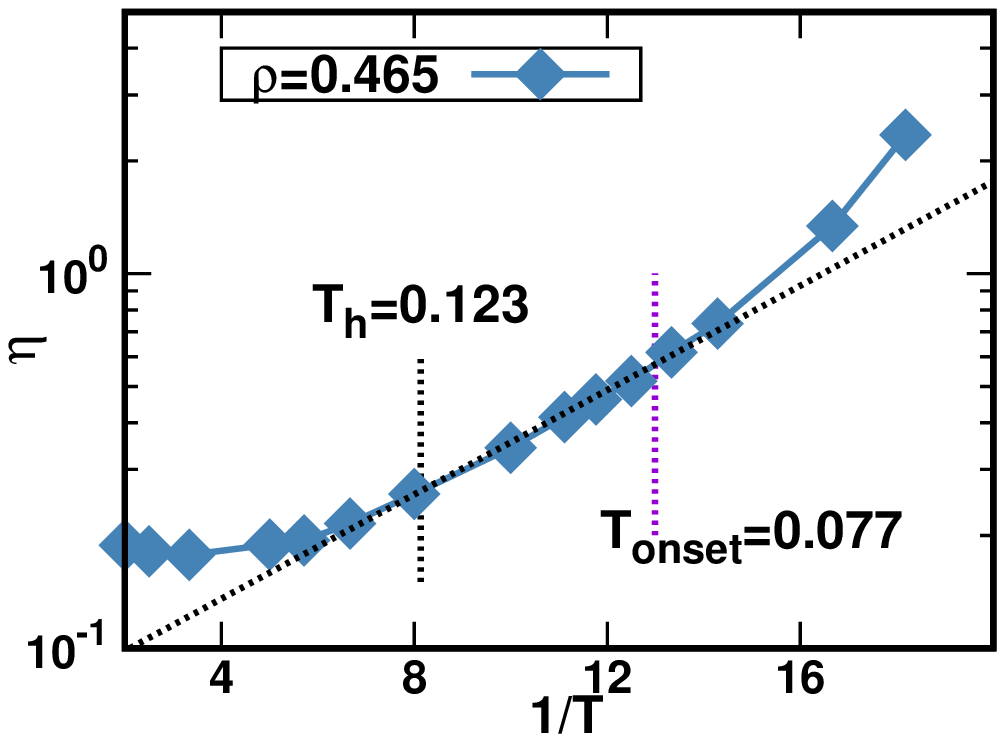}
    \includegraphics[width=0.3\textwidth]{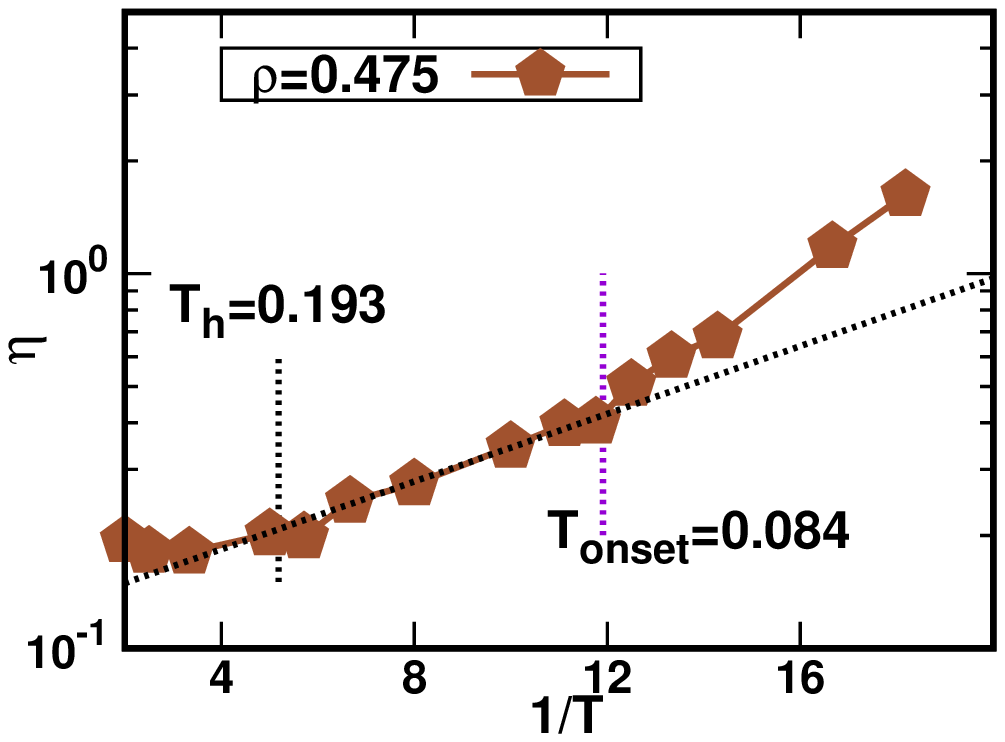}
    \put(-440,60){\textbf{(d)}}
    \put(-280,60){\textbf{(e)}}
    \put(-110,80){\textbf{(f)}}
    \caption{\emph{Timescales and characteristic temperatures along isochores of SW silicon.} We show the temperature ($T$) dependence of \textbf{(a)-(c)} diffusion coefficient ($D$), and \textbf{(d)-(f)} shear viscosity ($\eta$) for SW silicon along different isochores. Black dotted lines are fits to Arrhenius law. At low temperature, there is a crossover to non-Arrhenius behavior below the onset temperature $T_{onset}$. We also observe a high-temperature deviation from Arrhenius behavior above a temperature $T_h$. }
    \label{fig:TdepIsochore}
\end{figure*}

\begin{figure*}
    \centering
     \includegraphics[width=0.3\textwidth]{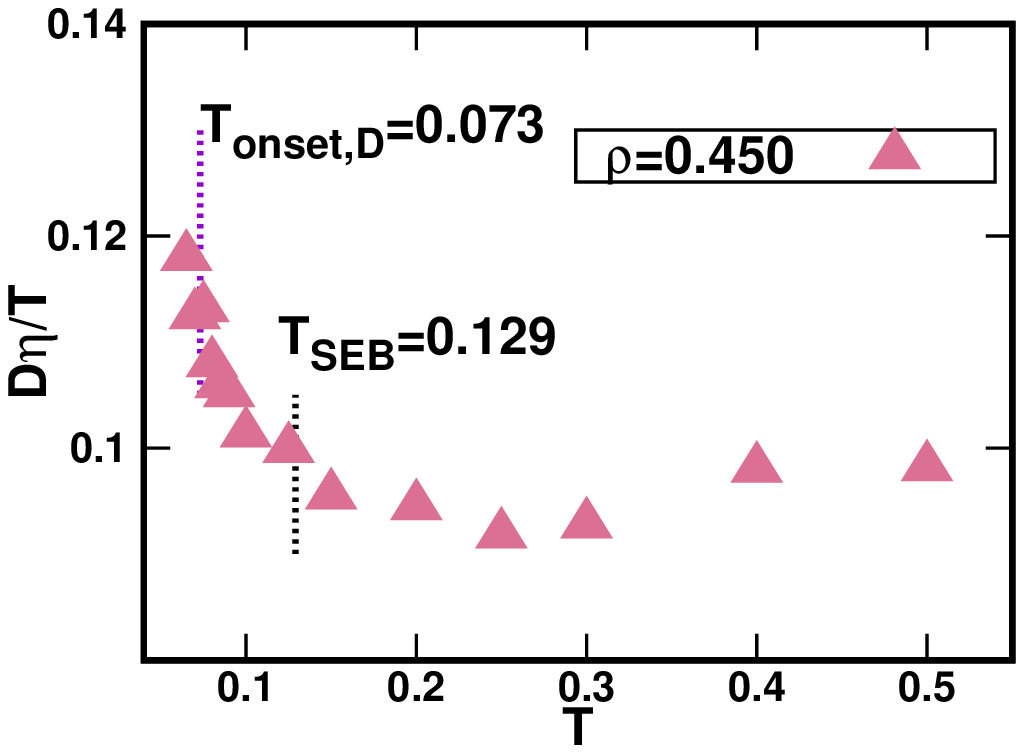}
     \includegraphics[width=0.3\textwidth]{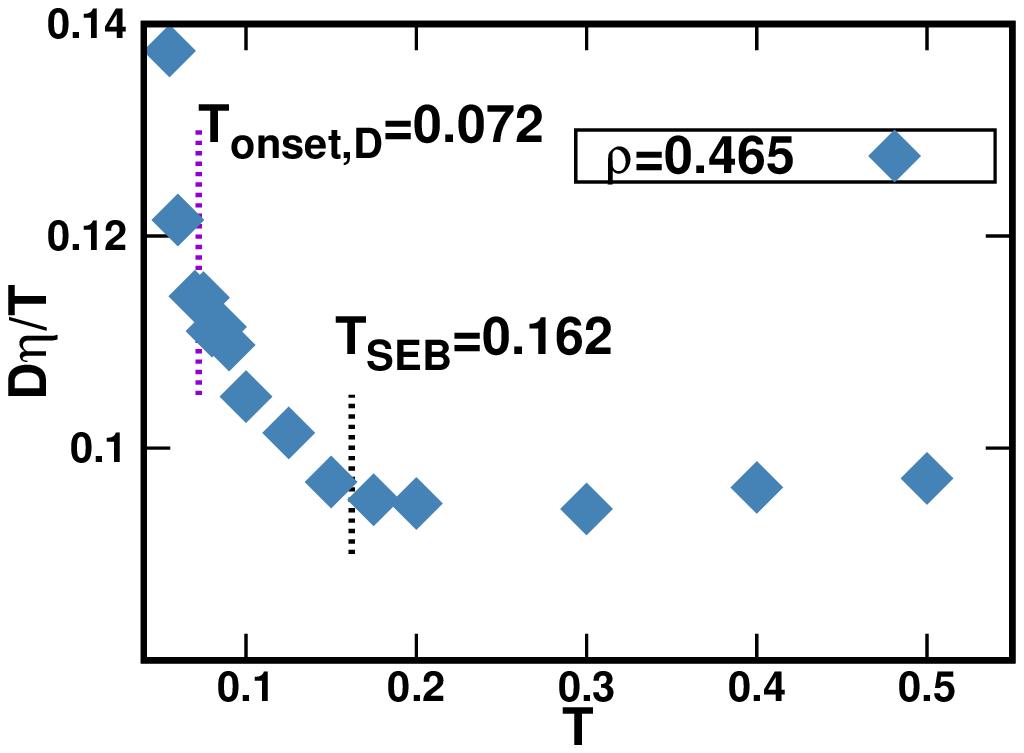}
     \includegraphics[width=0.3\textwidth]{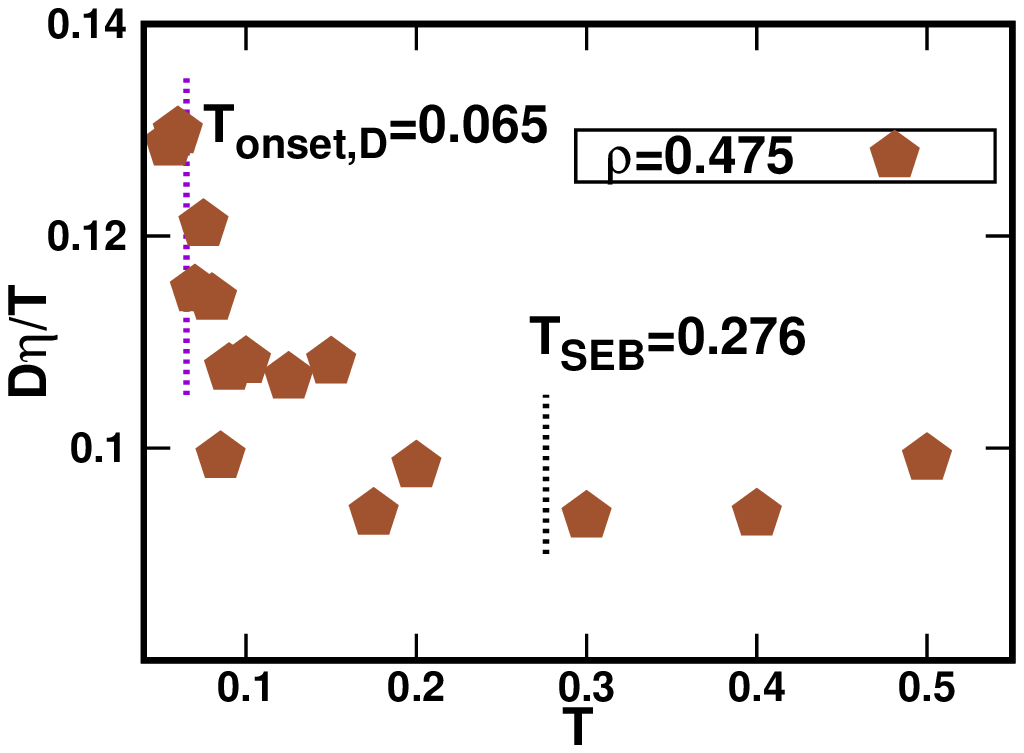}
     \put(-360,60){\textbf{(a)}}
     \put(-200,60){\textbf{(b)}}
     \put(-50,95){\textbf{(c)}}
     \\
     \includegraphics[width=0.3\textwidth]{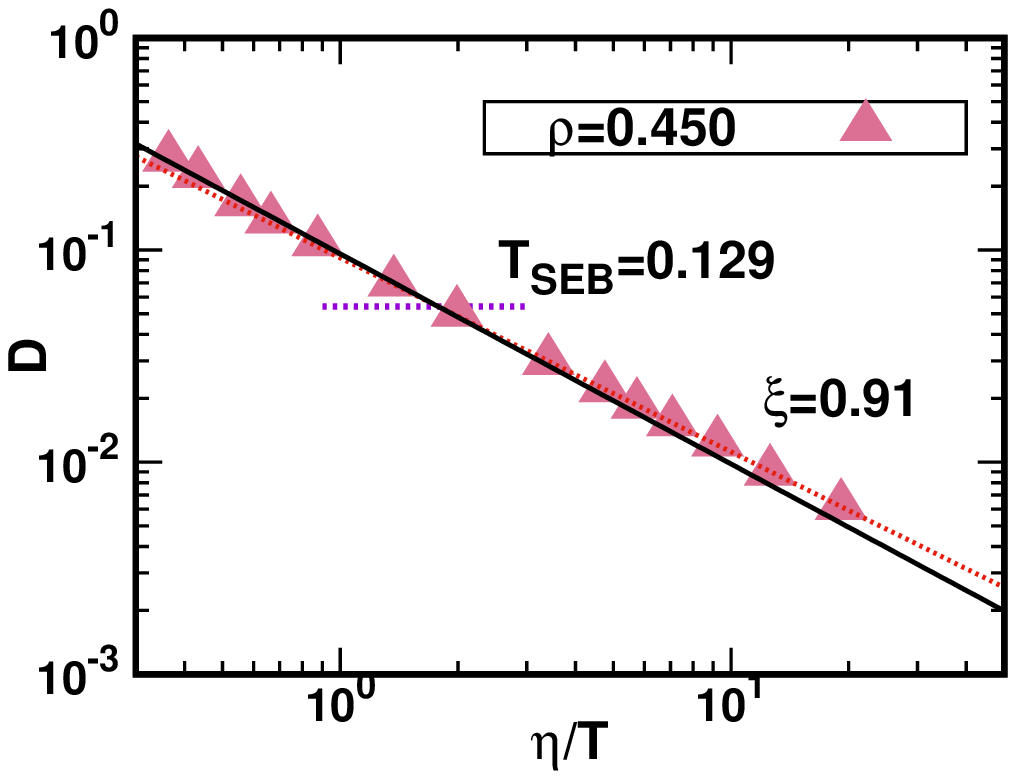}
     \includegraphics[width=0.3\textwidth]{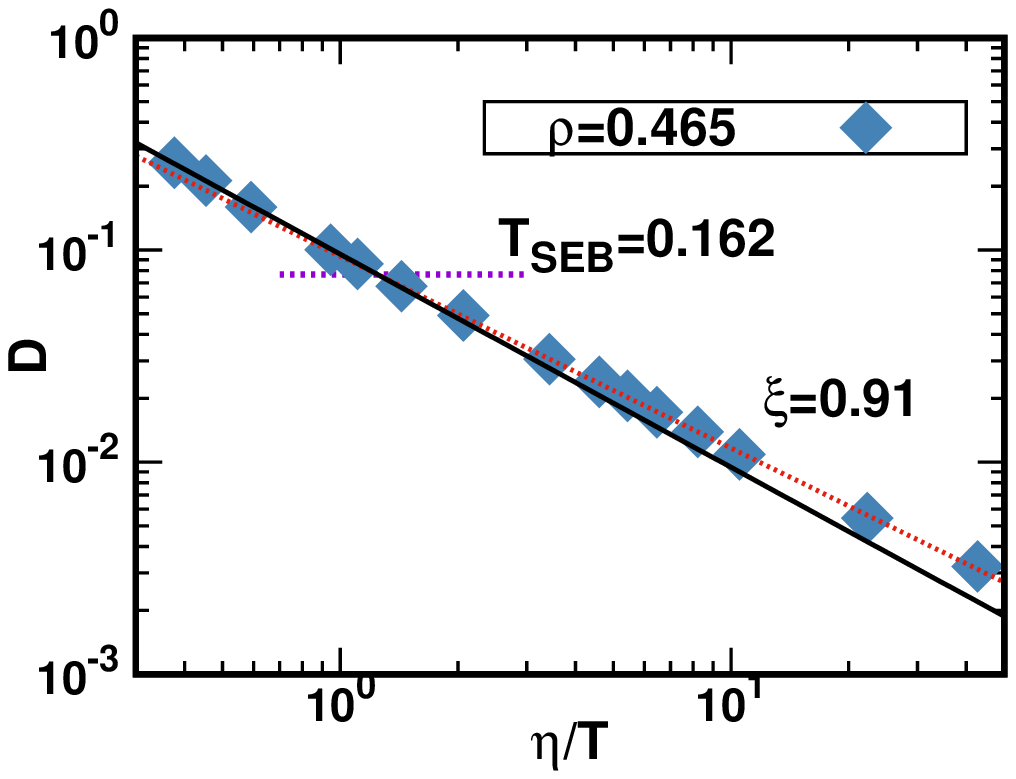}
     \includegraphics[width=0.3\textwidth]{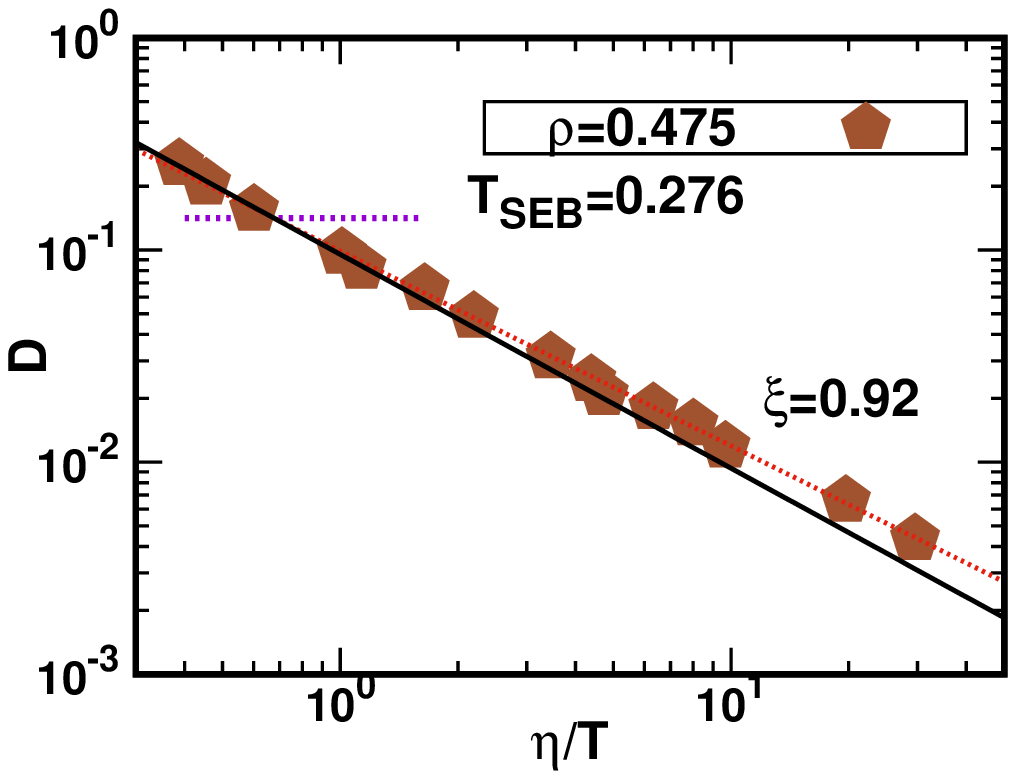}
     \put(-410,30){\textbf{(d)}}
     \put(-240,30){\textbf{(e)}}
     \put(-100,30){\textbf{(f)}}
     \caption{\emph{Testing the Stokes-Einstein relation along isochores of SW Silicon.} The Arrhenius regime is observed at higher temperatures along isochores, so the SER can be tested over a broader range of temperatures. \textbf{(a)-(c):} Temperature variation of the SE ratio $\frac{D\eta}{T}$. The \Revision{breakdown of the SE relation} (SEB) is observed at a temperature $T_{SEB}$ much higher than the onset temperature, see Table \ref{tab:isochore}. \textbf{(d)-(f):} Plots of $D$ {\it vs.} $\frac{\eta}{T}$, show that contrary to the case of isobars, along isochores SW silicon displays a normal SE regime deep into the Arrhenius regime, denoted by the solid fit line with $\xi=1$, Eqn. \ref{eqn:SER}. As $T$ decreases, there is a systematic deviation from the SER, described by dotted fit lines to fractional SE relation, $\xi < 1$, Eqn. \ref{eqn:fracSE}. We find $\xi \approx 0.9$ at all isochores analyzed, indicating the breakdown is weak and comparable to that along isobars. Note that the breakdown temperatures $T_{SEB} > T_{onset}$ {\it i.e.} lie in the Arrhenius regime.}
     \label{fig:SEBisochore}
\end{figure*}

\begin{figure*}[htbp]
  \centering
  \includegraphics[width=0.28\textwidth, height=5.0cm]{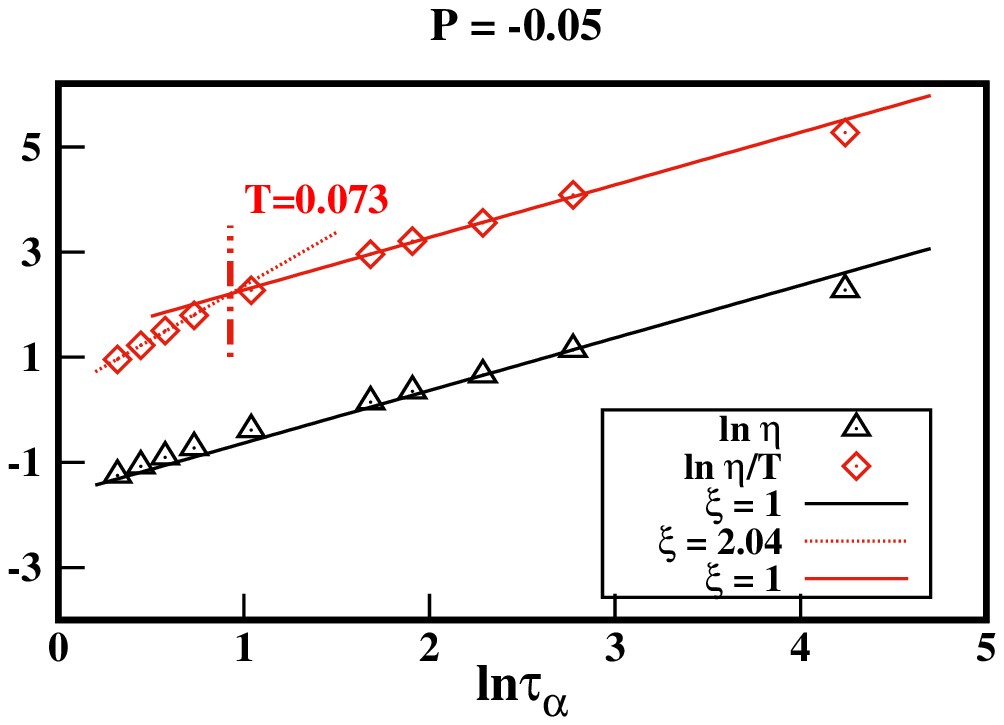}
  \includegraphics[width=0.28\textwidth, height=5.0cm]{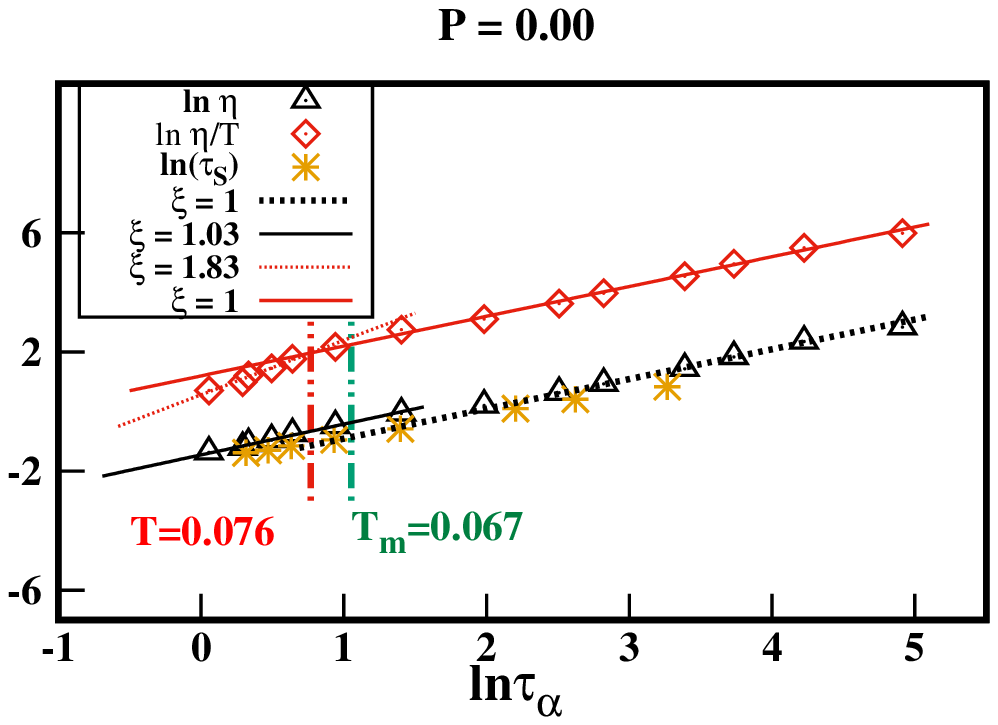}
  \includegraphics[width=0.28\textwidth, height=5.0cm]{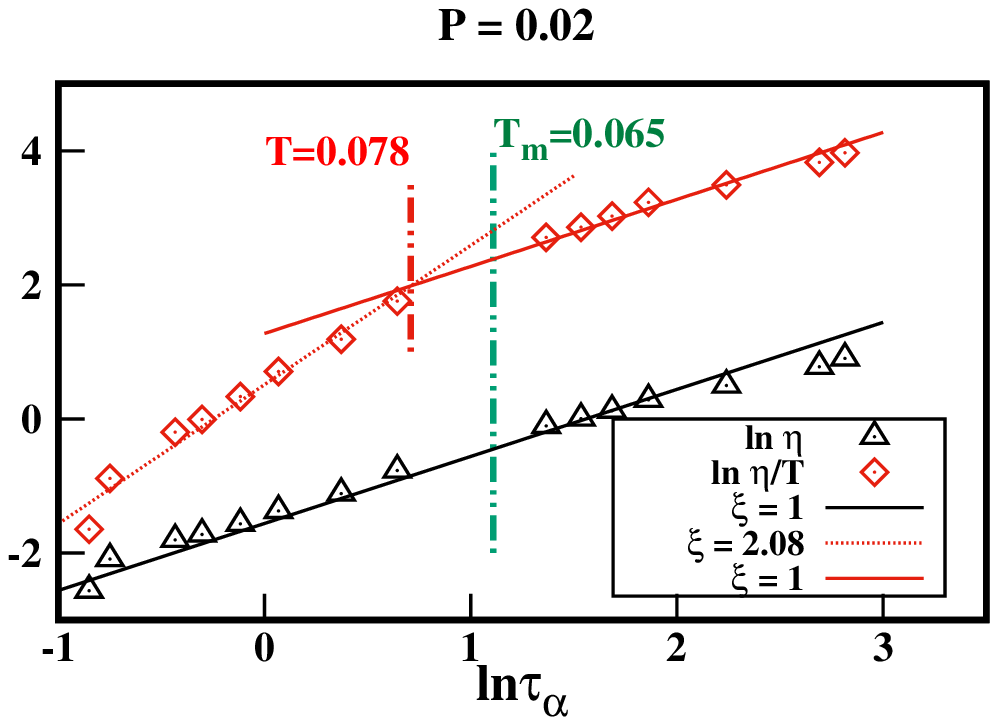}
  \put(-370,110){\textbf{(a)}}
  \put(-230,110){\textbf{(b)}}
  \put(-40,80){\textbf{(c)}}
  \\
  \includegraphics[width=5.0cm, height=4.8cm]{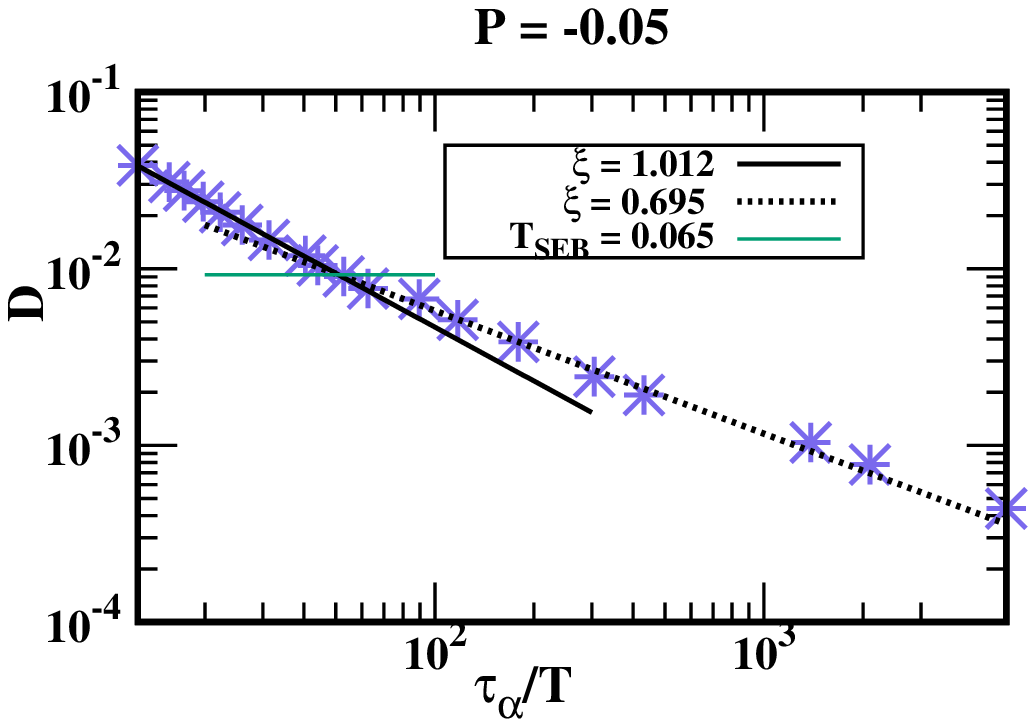}
  \includegraphics[width=5.0cm, height=4.8cm]{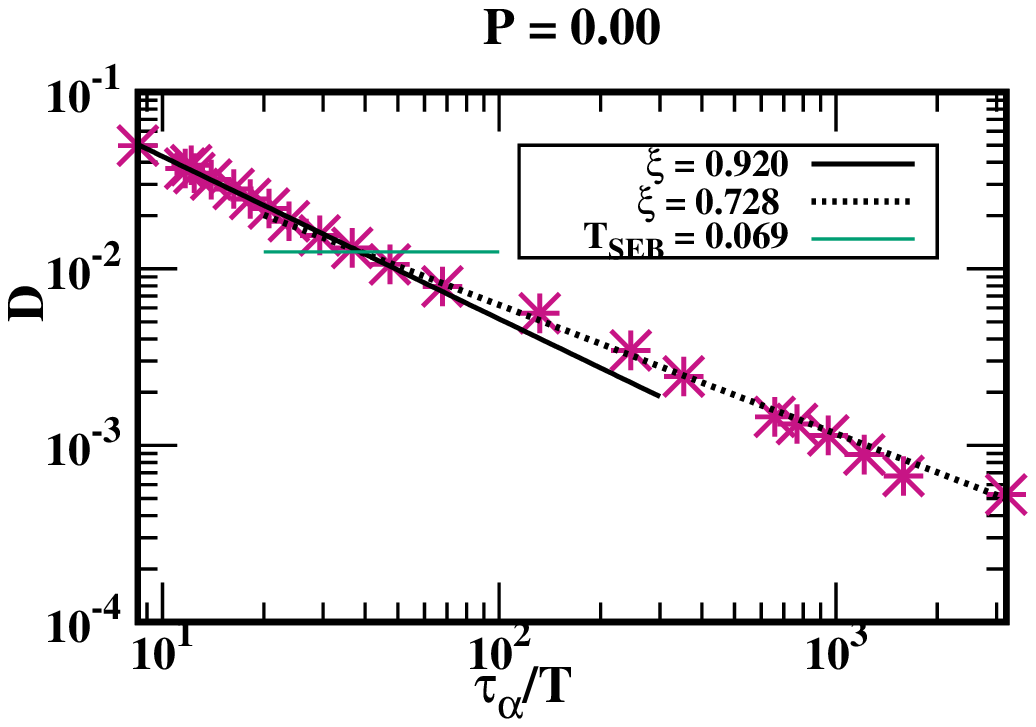}
  \includegraphics[width=5.0cm, height=4.8cm]{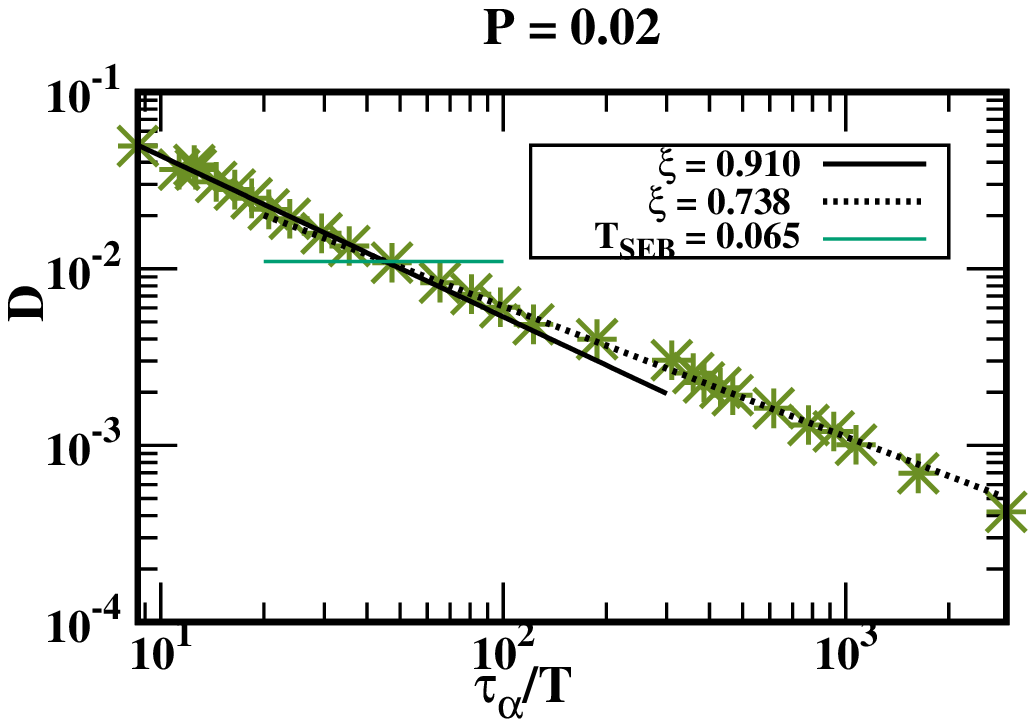}
  \put(-400,40){\textbf{(d)}}
  \put(-230,40){\textbf{(e)}}
  \put(-80,40){\textbf{(f)}}
  \\
  \includegraphics[width=5.0cm, height=4.8cm]{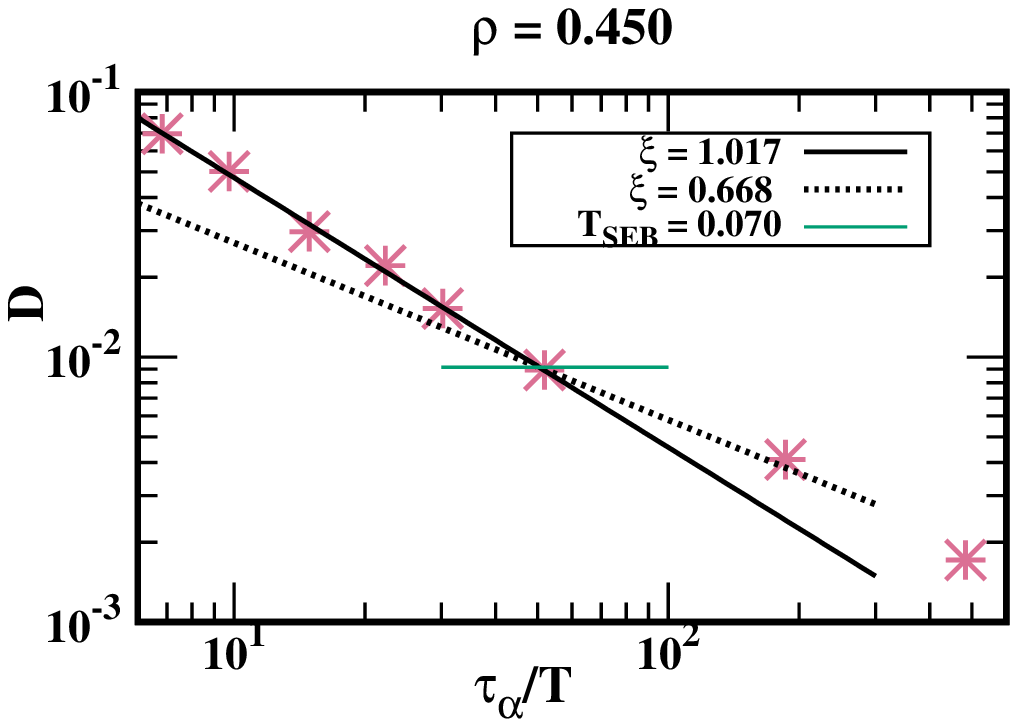}
  \includegraphics[width=5.0cm, height=4.8cm]{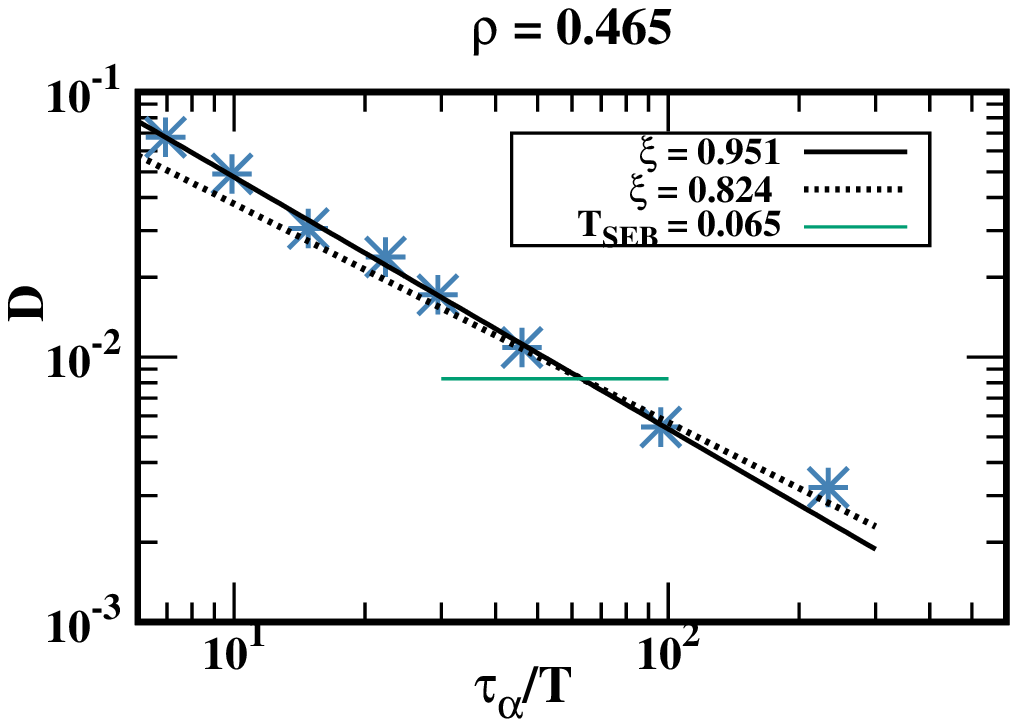}
  \includegraphics[width=5.0cm, height=4.8cm]{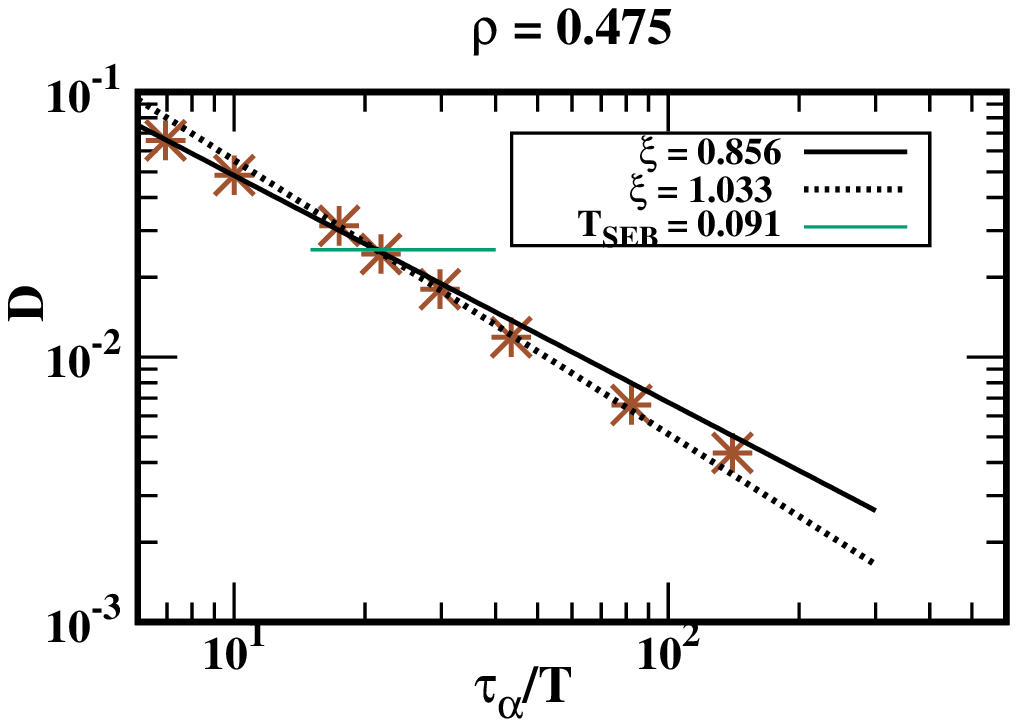}
  \put(-400,40){\textbf{(g)}}
  \put(-230,40){\textbf{(h)}}
  \put(-80,40){\textbf{(i)}}
  \caption{\emph{Testing SEB using $\tau_\alpha$.} 
    \textbf{(a)-(c)}: Comparison of $\eta$ and $\frac{\eta}{T}$ with $\tau_\alpha$ along representative negative \textbf{(a)}, zero \textbf{(b)} and positive \textbf{(c)} pressure isobars show that approximate proportionality of $\tau_\alpha$ and $\eta$  holds over a larger range of $T$ than that between $\tau_\alpha$ and $\frac{\eta}{T}$. In other words in SW silicon $\tau_\alpha$ is a better proxy to $\eta$ than $\frac{\eta}{T}$ over a broad range of pressures. Here $\tau_\alpha$ is measured at $q^*$ which is the first peak of static structure factor $S(q)$, see Fig. \ref{fig:SQ}. Along $P=0.0$ isobar, comparison between the stress relaxation timescale $\tau_s=\frac{\eta}{G_\infty}$ and $\tau_\alpha$ is also shown. $G_\infty$ is the instantaneous shear modulus, see Fig. \ref{fig:TdepIsobar}(e) for its $T$ dependence.
    \textbf{(d)-(i):} Testing the SER using $\tau_\alpha$ as a proxy for $\eta$ along representative negative \textbf{(d)}, zero \textbf{(e)} and positive \textbf{(f)} pressure isobars and along representative isochores [\textbf{(g)-(i)}]. We see that along both isobars and isochores, there are two distinct regimes with very different exponents and a clear cross-over. Since the high $T$ exponent is closer to 1 (except at the highest density $\rho=0.475$), we interpret the crossover temperature as $T_{SEB}$. 
  }
  \label{fig:etavstau}
\end{figure*}

\begin{figure}[htbp]
    \centering
    \includegraphics[width=0.32\textwidth]{SOFQ_RU_P0.00.eps}
    \put(-100,90){\textbf{(a)}}
    \\
    \includegraphics[width=0.32\textwidth]{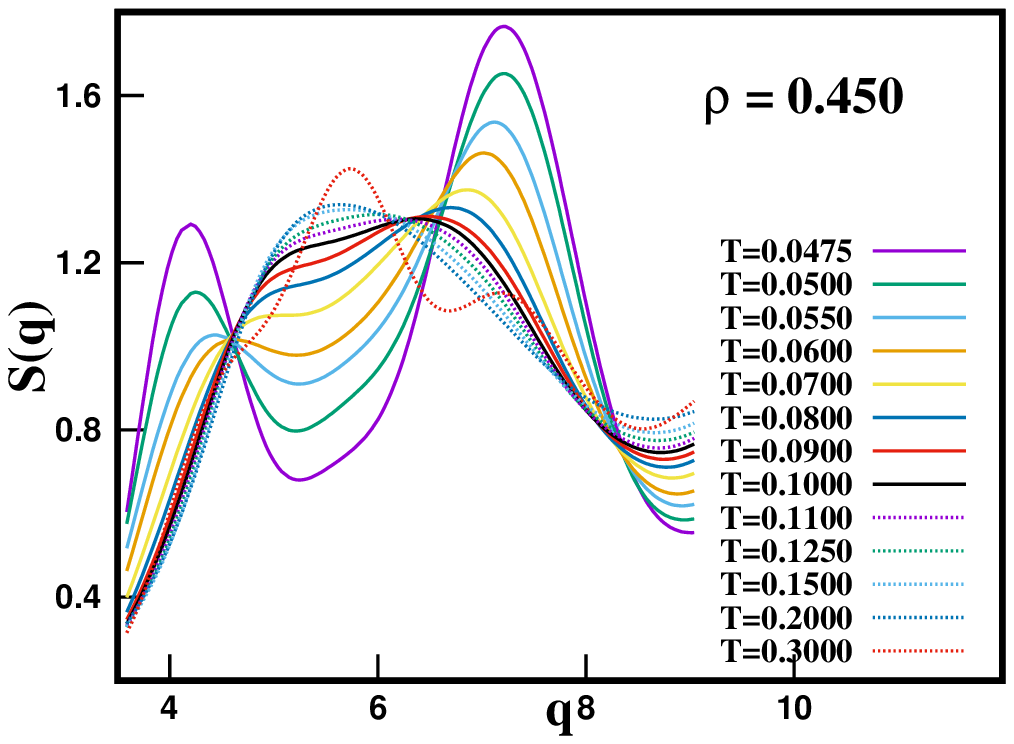}
    \put(-130,90){\textbf{(b)}}
    \caption{\textbf{(a):} Evolution of structure factor $S(q)$ with temperature ($T$) for $P=0.0$ isobar. (inset): $T$ dependence of the first peak position $q^{*}$. \Revision{\textbf{(b):}} $T$ dependence of $S(q)$ along a representative isochore for comparison. }
    \label{fig:SQ}
  \end{figure}

Among network-forming liquids, the SEB has been extensively documented in water \cite{Becker2006, Kumar2007, Mazza2007, Xu2009, KK2017, SRT2018}, but the behaviour of silicon is not so well studied. To characterize the dynamics of silicon, we consider three standard measures - (i) translational diffusion coefficient $D$, (ii) the shear viscosity $\eta$ and (iii) the $\alpha$-relaxation time $\tau_\alpha$. In Fig. \ref{fig:TdepIsobar} we show the temperature ($T$) dependence of $D$ [(a)-(c)], $\eta$ [(d)-(f)] and $\tau_\alpha$ [(g)-(i)] along isobars. For all measures of dynamics, we identify an Arrhenius regime shown by fit lines, and non-Arrhenius behaviour below the characteristic cross-over temperature $T_{onset}$ marking the onset of slow dynamics. Interestingly we also observe a deviation from Arrhenius behaviour also above a high temperature $T_h$. For the current work, we consider only temperatures below $T_h$ for further analyses. The values of the characteristic temperatures are listed in Table \ref{tab:isobar}. \Revision{Uncertainty in computed $T_{onset}$ value is estimated in Fig. \ref{fig:Tonset_errorbar} in the Appendix.} We note that in SW silicon, $T_{onset} < T_m$ along isobars. This is in contrast to liquids with isotropic interactions {\it e.g.} the 3d Kob-Andersen model, where $T_{onset} \approx T_m$ \cite{Sengupta2013, 2018Pedersen}.

Utilizing the measures $D$, $\eta$ and $\tau_\alpha$,  we test the validity of the SER along isobars of SW silicon in two different ways. In \ref{fig:SEBisobar}(a)-(c) we show the $T$ dependence of the SE ratio $\frac{D\eta}{T}$. In none of the isobars it reaches a constant value at high temperatures, thus violating Eqn. \ref{eqn:SER} over a wide range of temperatures varying from above to below $T_m$. To better understand the dependence, we show in Fig. \ref{fig:SEBisobar}(d)-(f) examines the power law relationship between $D$ and $\frac{\eta}{T}$, and we observe a fractional SE relation, Eqn. \ref{eqn:fracSE}, over the entire temperature range analyzed along all the isobars. The fractional exponent $\xi \approx 0.9$ does not depend significantly on pressure, see Table \ref{tab:isobar}. Thus, we find a weak SEB along isobars spanning a broad range from negative to positive pressures in SW silicon. In other words, even above $T_{onset}$ in the Arrhenius regime, $D$ and $\frac{\eta}{T}$ are (weakly) decoupled. 

\begin{table*}
  \centering
  \caption{Characteristic temperatures for SW Silicon along isobars. The onset temperatures are obtained from Fig. \ref{fig:TdepIsobar}, $T_{SEB}$ and the SE exponents are obtained from fits, see Fig. \ref{fig:SEBisobar} and \ref{fig:etavstau}. The suffix $h$ and $l$ denote high and low temperature regimes respectively.}
  \begin{tabular}{|c|c|c|c|c|c|c|c|c|c|c|c|}
    \hline
    $P$ & $T_{h,D}$ & $T_{onset,D}$  &  $T_{onset,\eta}$ & $T_{onset,\tau_\alpha}$ & $\xi_{\eta}$ & $T_{SEB,\tau_\alpha}$ & $\xi_{h,\tau_\alpha}$ & $\xi_{l,\tau_\alpha}$ \\
    \hline
    -0.05 & -      & 0.060         & 0.070            &  0.075                 & 0.88     & 0.065 & 1.01 & 0.70 \\
    -0.03 & 0.134  & 0.060         & 0.064            &  0.065                 & 0.89     & 0.071 & 0.91 & 0.78 \\
    0.00  & 0.130  & 0.053         & 0.060            &  0.065                 & 0.91     & 0.069 & 0.92 & 0.73 \\
    0.02  & 0.143  & 0.050         & 0.055            &  0.065                 & 0.90     & 0.065 & 0.91 & 0.74 \\
    \hline  
  \end{tabular}
  \label{tab:isobar}
\end{table*}

\begin{table*}
  \centering
  \caption{Characteristic temperatures for SW Silicon along isochores. The onset temperatures are obtained from Fig. \ref{fig:TdepIsochore}, $T_{SEB}$ and the SE exponents are obtained from fits, see Fig. \ref{fig:SEBisochore} and \ref{fig:etavstau}. The suffix $h$ and $l$ denote high and low temperature regimes respectively. }
  \begin{tabular}{|c|c|c|c|c|c|c|c|c|c|c|}
    \hline
    $\rho$ & $T_{h,D}$ & $T_{onset,D}$ & $T_{h,\eta}$ & $T_{onset,\eta}$ & $T_{SEB,\eta}$ & $\xi_{h,\eta}$ & $\xi_{l,\eta}$ & $T_{SEB,\tau_\alpha}$ & $\xi_{h,\tau_\alpha}$ & $\xi_{l,\tau_\alpha}$\\
    \hline
    0.450 &	0.364 & 0.073         & 0.133       & 0.091           & 0.129         & 0.99          & 0.91          & 0.070 & 1.02 & 0.67\\
    0.465 &	0.221 & 0.072         & 0.123	    & 0.077           & 0.162         & 1.01          & 0.91          & 0.065 & 0.95 & 0.82\\
    0.475 &     0.220 & 0.065         &	0.193       & 0.084           & 0.276         & 1.01          & 0.92          & 0.091 & 0.86 & 1.03\\
    \hline
  \end{tabular}
  \label{tab:isochore}
\end{table*}

\subsubsection{$T_{SEB} > T_{onset}$ along isochores}
The question naturally arises whether the normal SE regime is observed for the SW silicon. Fig. \ref{fig:phaseDiag_SWSi} shows that a broader range of liquid state points of silicon can be accessed along isochores. In Fig. \ref{fig:TdepIsochore} we show the $T$ dependence of $D$ (panels \ref{fig:TdepIsochore}(a)-(c)) and $\frac{\eta}{T}$ (panels \ref{fig:TdepIsochore}(d)-(f)) for three isochores and tabulate the characteristic temperatures in Table \ref{tab:isochore}. The values of $T_{onset}, T_h$ indicate that the Arrhenius regime along isochores indeed occur at higher temperatures than along isochores. Note that the melting line is not encountered along the isochores studied here, see Fig. \ref{fig:phaseDiag_SWSi}.

Fig. \ref{fig:SEBisochore} tests the validity of the SER along isochores. Panels \ref{fig:SEBisochore}((a)-(c)) describe the $T$ dependence of the SE ratio $\frac{D\eta}{T}$ along various isochores while Figs. \ref{fig:SEBisochore}((d)-(f)) compares $D$ {\it vs.} $\frac{\eta}{T}$. The main observation is that along isochores the SER, Eqn \ref{eqn:SER}, is obeyed at high temperatures with an exponent $\xi_h \approx 1$. A fractional SER, Eqn. \ref{eqn:fracSE}, with exponent $\xi_l \sim 0.9$, is followed below a characteristic temperature $T_{SEB}$. Thus the data indicates a weak breakdown of the SE relation in SW silicon along isochores. We emphasize that $T_{SEB}$ values are found to be much higher than $T_{onset}$ and are located in the Arrhenius regime (except at the highest density where it is  $> T_h$).

\subsubsection{SEB using $\tau_\alpha$}\label{sec:SEBtau}
To gain insight into the structural relaxation processes, one often substitutes $\eta$ by the $\alpha$-relaxation time $\tau_\alpha$ in the SER, Eqn. \ref{eqn:SER} \cite{Ediger2000, 2006Chen, 2007Alonso, Xu2009}. However, unlike Eqn. \ref{eqn:SER}, there is no direct theoretical relation between $\eta$ and $\tau_\alpha$. The Maxwell's model suggests that $\eta$ is related to the \emph{stress} relaxation time $\tau_{s}$ as $\eta = G_\infty \tau_{s}$, where $G_\infty$ is the instantaneous shear modulus. $\tau_{s}$ is in turn considered to be proportional to $\tau_\alpha \equiv \tau(q^*)$, obtained from \emph{density} relaxation. \Revision{Although we use $G_\infty$ in the present analysis, $\eta$ has also been related to elastic modulus at longer times {\it e.g.} plateau modulus $G_p$ \cite{Puosi2012, KK2017}. It is an interesting issue that requires further study.} However, in the absence of a rigorous derivation, it is not {\it a priori} clear whether $\tau_\alpha$ should be a proxy for $\eta$ or $\frac{\eta}{T}$ \cite{Shi2013}, {\it i.e.} whether the difference in the factor of $T$ is significant or not.  

Thus, we first test the proportionality of both $\eta$ and $\frac{\eta}{T}$ with $\tau_\alpha$ over a broad regime of the phase diagram of SW silicon. In Fig. \ref{fig:etavstau}(a)-(c), we show the comparison for different isobars along with power law fits. The exponent $\xi = 1$ indicates perfect coupling. For the zero pressure case, both $\eta$ and $\frac{\eta}{T}$ show good coupling to $\tau_\alpha$ at \emph{low} $T$. Interestingly their differences are marked at \emph{high} $T$, where exponents are close to 1 only for $\eta$. Similar behaviour is also seen along other isobars. Note that as shown in Fig. \ref{fig:SQ} the temperature evolution of structure factor $S(q)$ is significant and consequently $q^*$ itself is varying with temperature range studied here, inset of Fig. \ref{fig:SQ}(a), complicating the analysis. In Fig. \ref{fig:etavstau}(b), we also show the $T$ dependence of $\tau_s = \frac{\eta}{G_\infty}$ which can be described by a single power law regime. At zero pressure, $G_\infty$ is $\mathcal{O}(1)$ but shows a weak but clear $T$ dependence in the temperature range studied, see Fig. \ref{fig:TdepIsobar}(e). Hence for SW silicon although $\tau_s \approx \eta$ numerically, $\tau_s \propto \tau_\alpha$ is still an approximate relation. Nevertheless, taken together, the data suggest that $\tau_\alpha$ is a better proxy to $\eta$ than $\frac{\eta}{T}$ over larger span of state points in SW silicon.

We now test the SER using $D$ and $\frac{\tau_\alpha}{T}$ along isobars [Fig. \ref{fig:etavstau}(d)-(f)] and isochores [Fig. \ref{fig:etavstau}(g)-(i)]. In both cases, two different power-law regimes can be seen clearly. The high $T$ exponent is close to 1 at the lowest pressure (density) reported in this work, but as pressure (density) increases, the exponent value decreases. The low $T$ exponent however, shows the opposite dependence on pressure (density) and is more sensitive to change in density. In fact, for the highest density, the \emph{low} $T$ exponent is $\approx 1$. Nevertheless, we interpret the cross-over temperature between the two regimes as the \Revision{SEB} temperature $T_{SEB,\tau_\alpha}$ and report them in Tables \ref{tab:isobar} and \ref{tab:isochore}.

\begin{figure}[htbp]
    \centering
    \includegraphics[width=0.45\textwidth]{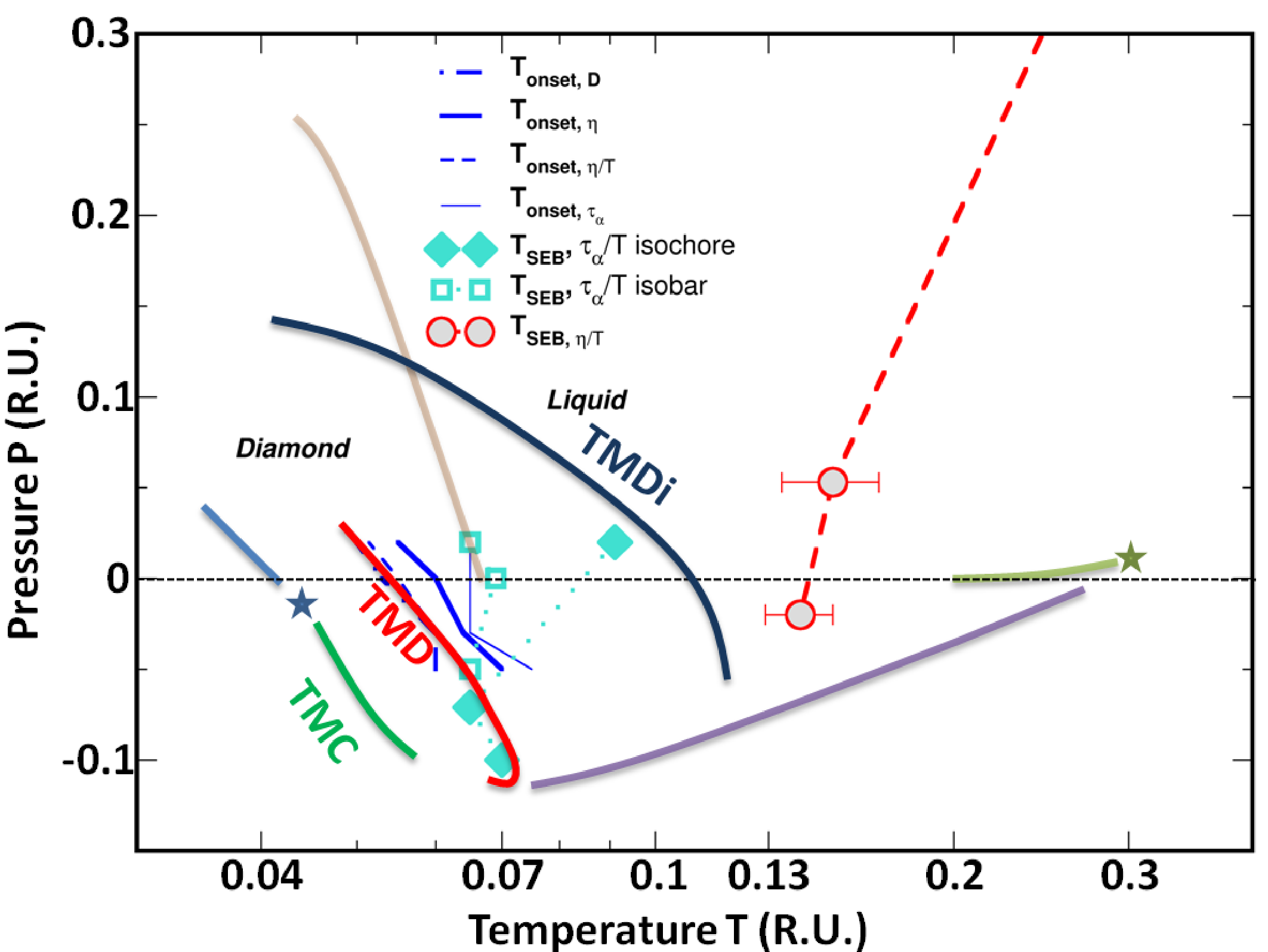}
    \caption{\emph{Loci of characteristic dynamical temperatures in the (P,T) plane for the SW silicon.} The values of the onset temperature $T_{onset}$ and the \Revision{SEB} temperature $T_{SEB}$ determined along isobars and isochores in the present study are shown by symbols. Various other characteristic thermodynamic, dynamic and structural features are shown by solid lines and are same as in Fig. \ref{fig:phaseDiag_SWSi}, \Revision{and are schematic adaptations from Refs. N. Phys. 7, 549 (2011) \cite{Vasisht2011}, Copyright 2011, Springer Nature Limited and ``Liquid Polymorphism: Advances in Chemical Physics'', Vol. 152 (2013), Ed. H. Eugene Stanley \cite{Vasisht2013}, Copyright 2013 John Wiley \& Sons, Inc.}}
    \label{fig:phaseDiag_loci}
\end{figure}

\begin{figure*}[htbp]
    \centering
    \includegraphics[width=0.32\textwidth]{SW-P0.00-Fkt-expon.eps}
    \includegraphics[width=0.32\textwidth]{SW-P0.00-Fkt-stretched.eps}
    \includegraphics[width=0.32\textwidth]{SW-P0.00-Fkt-twostep.eps}
    \put(-370,90){\textbf{(a)}}
    \put(-300,90){\textbf{(b)}}
    \put(-40,90){\textbf{(c)}}
    \\\vspace{3mm}
    \includegraphics[width=0.32\textwidth]{SW-P0.00-betaKWW.eps}\hspace{2mm}
    \includegraphics[width=0.32\textwidth]{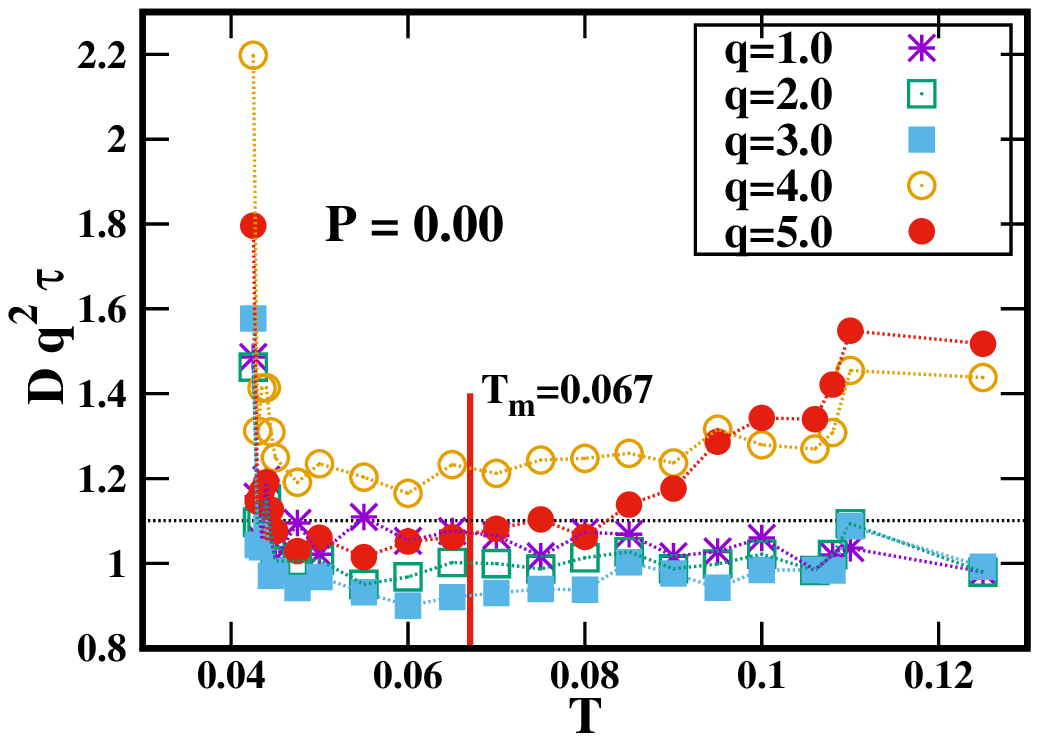}
    \includegraphics[width=0.32\textwidth]{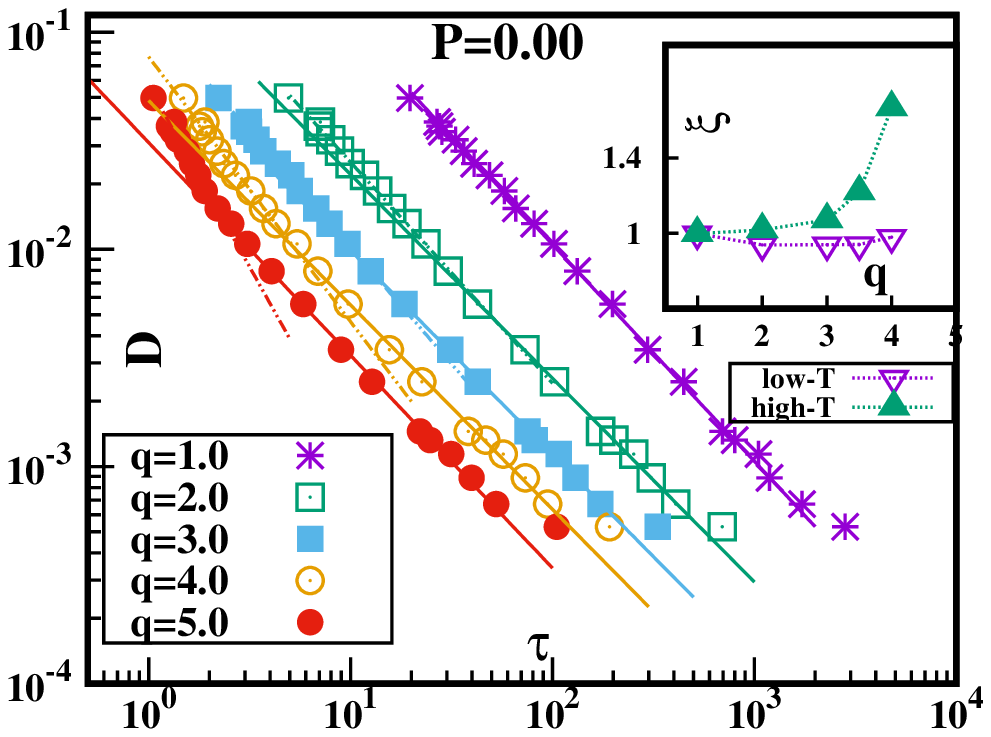}
    \put(-440,30){\textbf{(d)}}
    \put(-240,90){\textbf{(e)}}
    \put(-30,20){\textbf{(f)}}
    \caption{\emph{Arrhenius but non-Fickian regime along $P=0$ isobar.}
      \textbf{(a)-(c):} Intermediate scattering function $F(q,t)$ at representative $T$ and $q$ showing three types of decay profiles: \textbf{(a)} exponential decay as $q \rightarrow 0$; \textbf{(c)} typical two-step, \emph{non}-exponential decay characteristic of glassy dynamics as $q \rightarrow q^{*}$, in the supercooled regime, and \textbf{(b)} a \emph{single} step non-exponential decay in the intermediate $T$ and $q$ between these two limiting  cases. State points above and below $T_{onset,D}$ are marked in red and blue respectively. Solid lines are fits to (a) 1-parameter exponential $F(q,t) = e^{-\frac{t}{\tau_\alpha}}$ (b) 3-parameter, single-step  non-exponential $F(q,t) = f_c \, e^{-\left(\frac{t}{\tau_\alpha}\right)^{\beta_{KWW}}}$ and (c) 4-parameter, two-step non-exponential $F(q,t) = (1-f_c)\, e^{-\left(\frac{t}{\tau_s}\right)^2} + f_c\, e^{-\left(\frac{t}{\tau_\alpha}\right)^{\beta_{KWW}}}$ functions. Here $f_c$, denotes the plateau height, $\tau_s$ and $\tau_\alpha$ are relaxation times of short and long-time decay and the stretched exponent $\beta_{KWW}$ characterizes the nature and extent of deviation from exponential decay.
      \textbf{(d):} $\beta_{KWW}$ as a function of $T$ for different $q$ from low upto $\sim q^{*}$, quantifying the three types of decay profiles.
      \textbf{(e):} $T$ dependence of $Dq^2\tau(q)$ shows deviation from constancy - Eqn. \ref{eqn:Fickian} - in the \emph{high} $T$ \emph{Arrhenius} regime as $q \rightarrow q^{*}$, suggesting that the dynamics has not become Fickian yet at these timescales.
      \textbf{(f)} Testing the power law relationship between $D$ and $\tau(q)$ ($\tau(q)$ defined as $F(q,\tau)=1/e$). We see two different regimes with differing exponents. Inset shows the $q$ dependence of high and low temperature exponents suggesting that $D$ and $\tau (q)$ are apparently coupled at low temperature and decoupled at high temperature, in contradiction to existing phenomenology.}
    \label{fig:RawdataTimeP0}
\end{figure*}

  In Fig. \ref{fig:phaseDiag_loci}, we show the loci of the onset temperature $T_{onset}$ which characterizes a qualitative change from normal to activated dynamics. We compute $T_{onset}$ from several characteristic timescales - $D$, $\eta$ (also $\frac{\eta}{T}$) and $\tau_\alpha$. Although the precise numerical values vary depending on the quantity, we see that as pressure increases, $T_{onset}$ decreases. This signifies that in the anomalous region, the dynamics is faster at higher pressure, consistent with previous reports \cite{2017Handle}. More interestingly, we find the $T_{onset}$, in particular obtained from $D$ and $\frac{\eta}{T}$ along isobars, closely parallels the ``temperature of density maxima'' (TMD) line, obtained independently from earlier work \cite{Vasisht2011, Vasisht2013, Vasisht2014}, unlike water where the loci SEB coincides with the TMC or Widom line \cite{Kumar2007}.

  The behavior of the second characteristic temperature {\it i.e.} the $T_{SEB}$, however, depends on the choice of timescales. While the de-coupling of $D$ and $\tau_\alpha$ occur inside the region enclosed by lines of anomalies, and close to the melting line for non-negative pressures, the weak decoupling of $D$ and $\eta$ occur far away from the anomalous region.  This implies that anomalies influence different timescales differently. We mention that recent works on water based on two-state models \cite{SRT2018, 2020Shi} have predicted the occurrence of $T_{SEB}$ at relatively higher temperature, consistent with our findings in silicon. More importantly the loci of $T_{onset}$ and $T_{SEB}$ on the P-T plane, Fig. \ref{fig:phaseDiag_loci}, suggests the intriguing possibility of an in-between region where the dynamics is \emph{Arrhenius but non-Fickian}. In order to understand it in more detail, we analyze the dynamics at different lengthscales by varying the probe wavenumber $q$ in Sec. \ref{sec:qdep}.


\section{$q$-dependent dynamics} \label{sec:qdep}

\subsubsection{Fickian to non-Fickian crossover: Arrhenius but non-Fickian regime in SW silicon}

\begin{figure*}[htbp]
    \centering
    \includegraphics[width=0.32\textwidth]{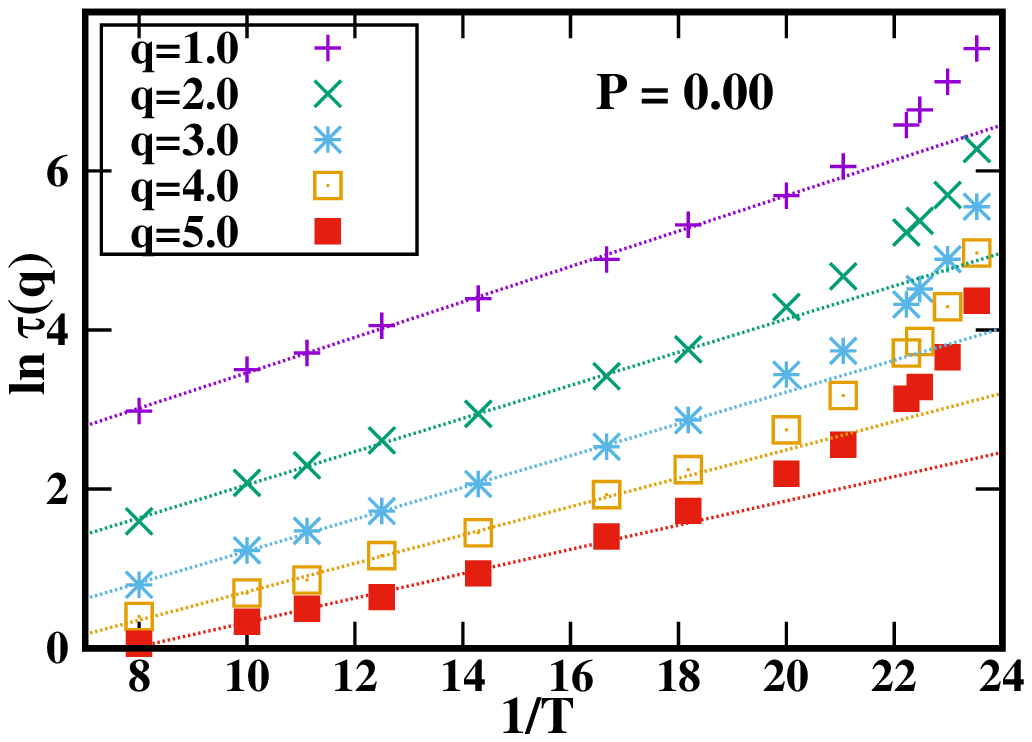}
    \includegraphics[width=0.32\textwidth]{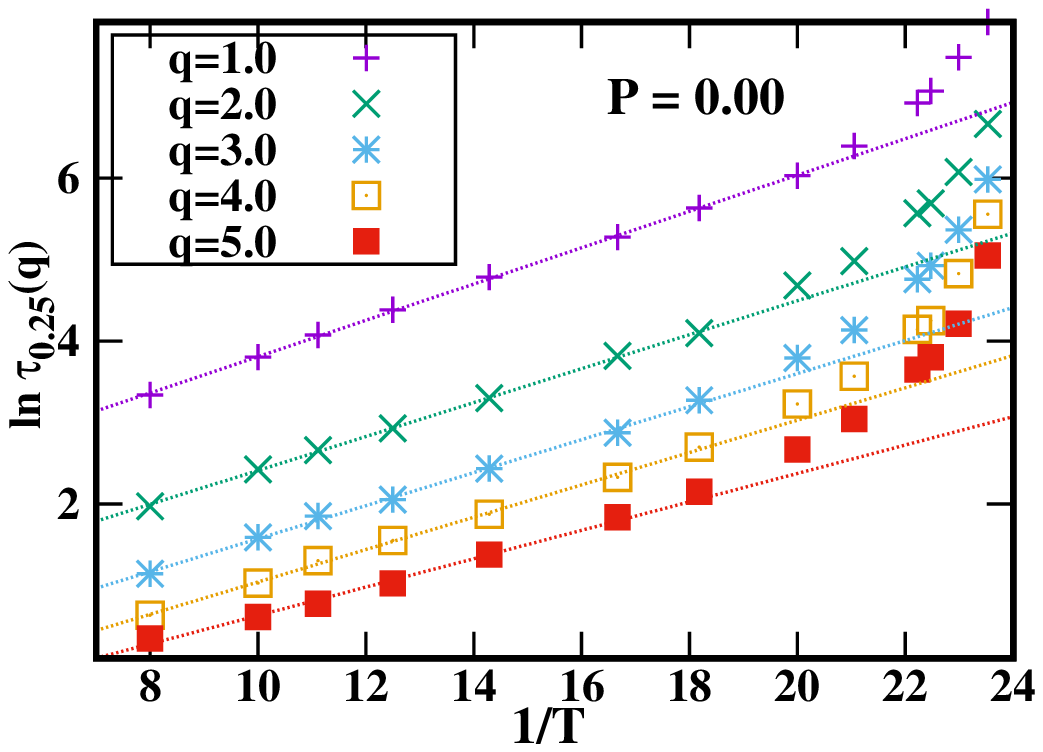}
    \includegraphics[width=0.32\textwidth]{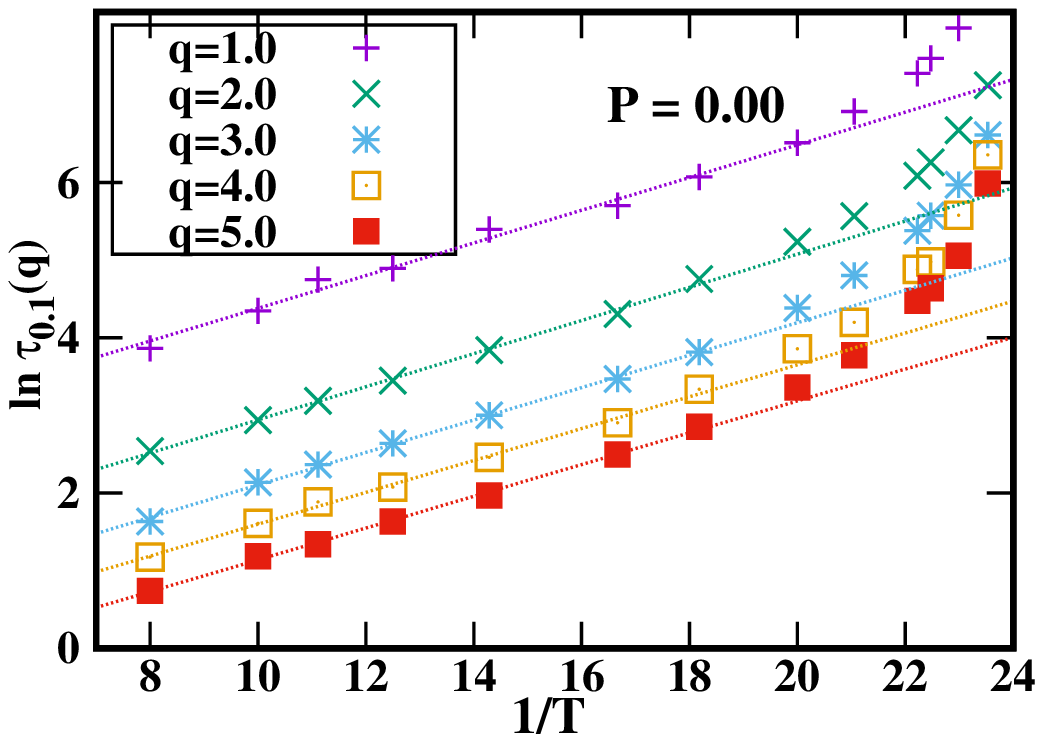}
    \put(-430,90){\textbf{(a)}}
    \put(-255,90){\textbf{(b)}}
    \put(-50,30){\textbf{(c)}}
    \\
    \includegraphics[width=0.35\textwidth]{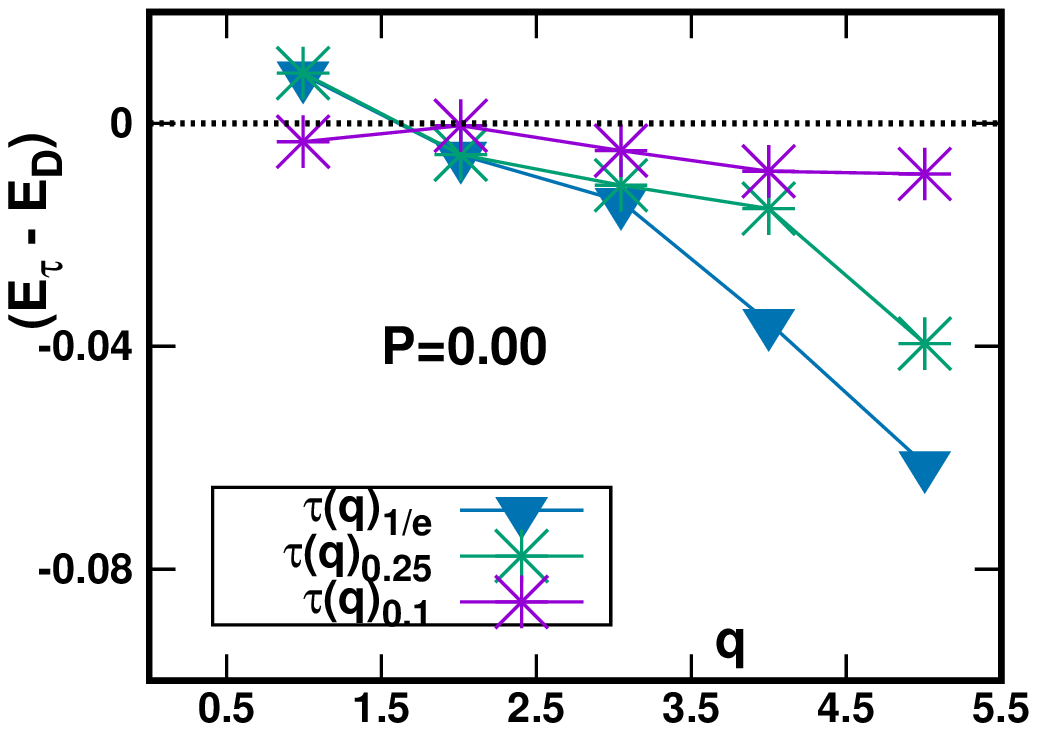}
    \includegraphics[width=0.35\textwidth]{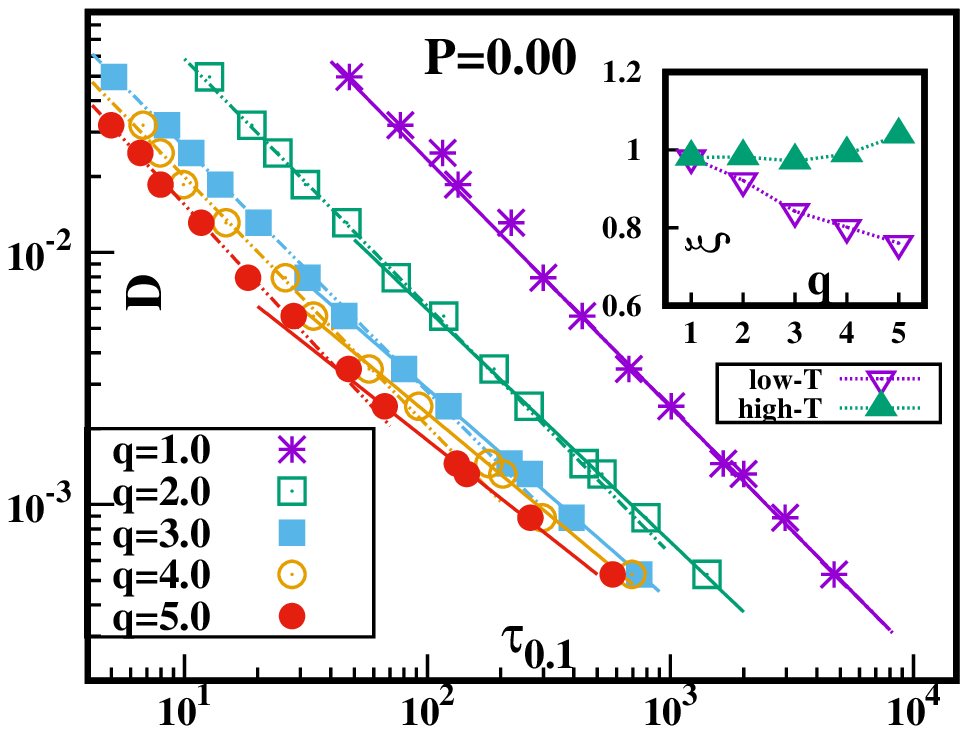}
    \put(-310,80){\textbf{(d)}}
    \put(-140,55){\textbf{(e)}}
    \caption{\emph{Effect of increasing timescales on dynamics:}
      \textbf{(a)-(c):} $T$ dependence of $q$-dependent relaxation time using the standard definition \textbf{(a)} $\tau_{1/e}$ and longer timescales \textbf{(b)} $\tau_{0.25}$ and \textbf{(c)} $\tau_{0.1}$ respectively, see Sec. \ref{sec:method} for definitions. 
      \textbf{(d):} We define the difference in activation energy as $\Delta E(q) = E_\tau(q) - E_D$ and analyze its $q$ dependence. We find significant $q$ dependence for the conventional measure of relaxation timescale, $\tau_{1/e}$, implying non-Fickian dynamics even at high $T$. However, as the probe timescale is gradually increased from $\tau_{1/e} \rightarrow \tau_{0.25} \rightarrow \tau_{0.1}$, the $q$ dependence $\Delta E(q)$ becomes weaker and approaches a constant approximately zero. It suggests that in SW silicon, true Fickian dynamics sets in at longer than conventional $\alpha$-relaxation timescale, \emph{even in the Arrhenius regime}.
      \textbf{(e):} Coupling and decoupling of $\tau_{0.1}(q)$  to $D$ {\it via} power law is shown in the main panel. Inset shows the $q$ dependence of exponents. The high T exponent is $\approx 1$ and low T exponent is $<1$ consistent with the Fickian picture of dynamics, Eqn. \ref{eqn:Fickian}.
    }
    \label{fig:qdepEnergy}
\end{figure*}

\begin{figure*}[htbp]
    \centering
    \includegraphics[width=0.3\textwidth]{SW-P0.00-nonFick-D-tau_inve.eps}\hspace{5mm}
    \includegraphics[width=0.3\textwidth]{SW-P0.00-nonFick-D-tau_0.25.eps}\hspace{3mm}
    \includegraphics[width=0.3\textwidth]{SW-P0.00-nonFick-D-tau_0.1.eps}
    \put(-420,90){\textbf{(a)}}
    \put(-250,90){\textbf{(b)}}
    \put(-80,90){\textbf{(c)}}
    \\\vspace{5mm}
    \includegraphics[width=0.28\textwidth, height=4cm]{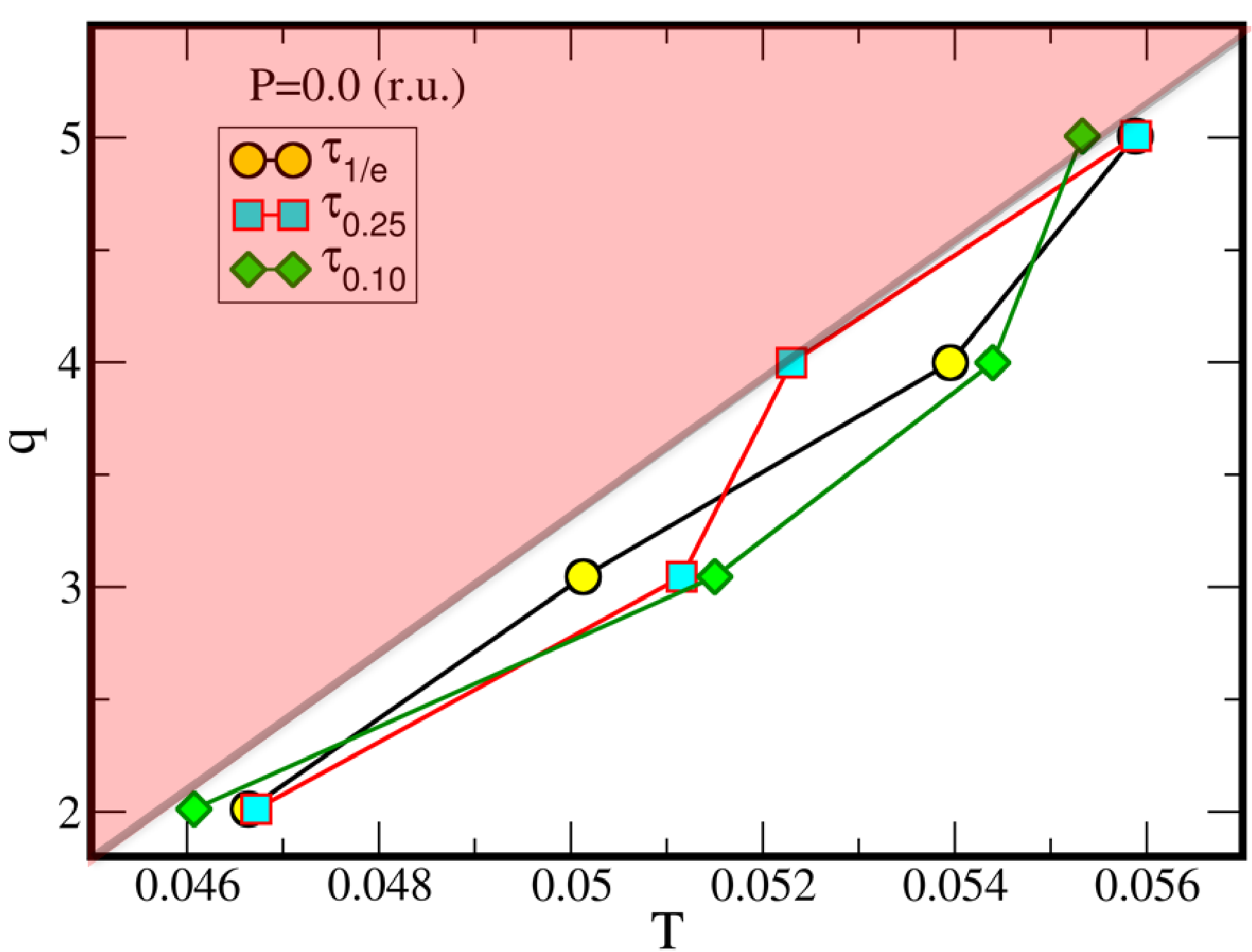}\hspace{5mm}
    \includegraphics[width=5.2cm, height=4cm]{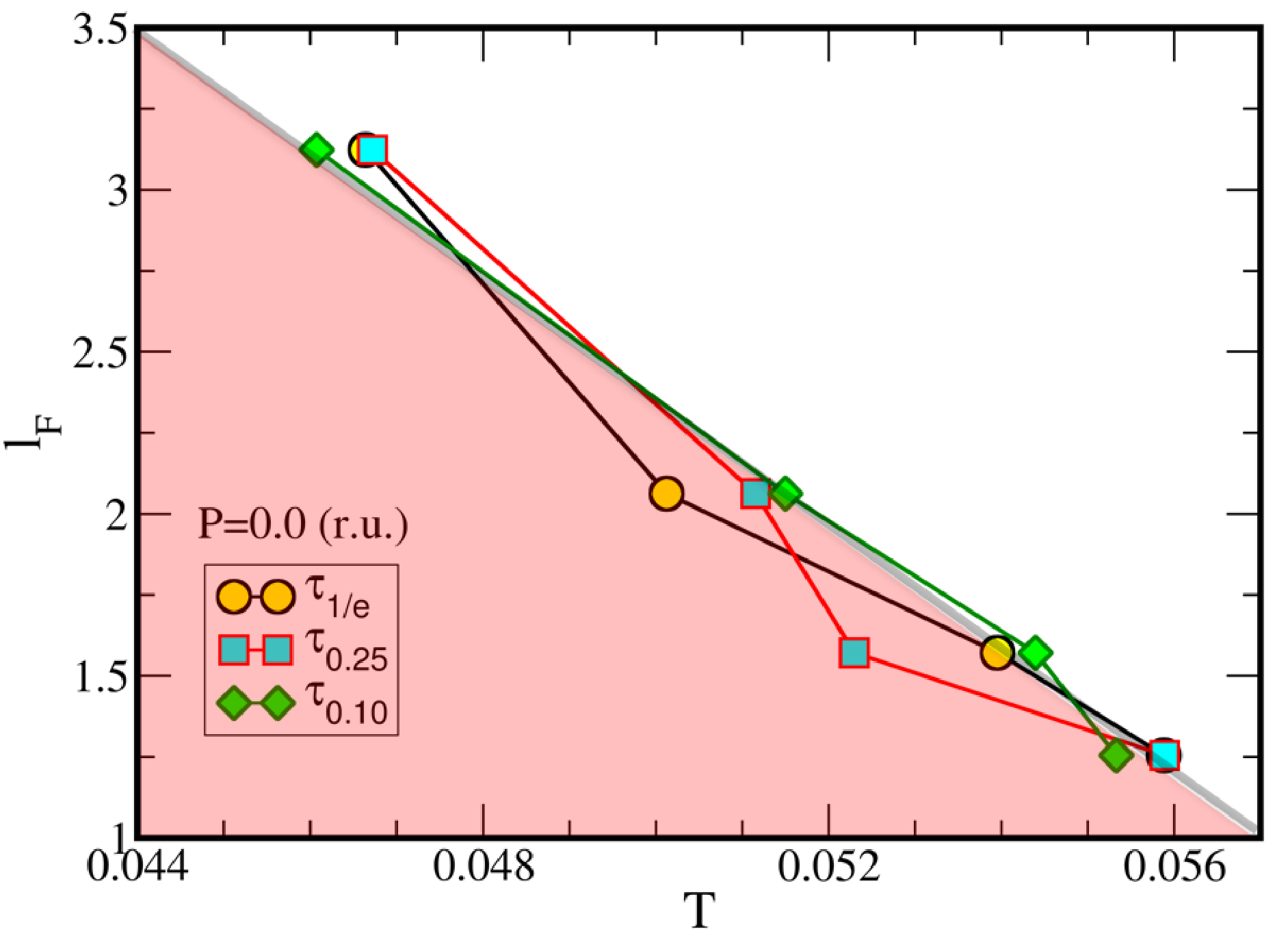}
    \put(-230,95){\textbf{(d)}}
    \put(-280,60){\textbf{SEB}}
    \put(-200,30){\textbf{SER}}
    \put(-70,95){\textbf{(e)}}
    \put(-90,30){\textbf{SEB}}
    \put(-60,70){\textbf{SER}}
    \\
    \caption{\emph{Fickian to non-Fickian cross-over lengthscale along $P=0$ isobar.}
      \textbf{\Revision{(a)-(c):}} $y \equiv \frac{D\tau}{D_0 \tau_0 \exp(\frac{\Delta E}{k_B T})}$ {\it vs.} $x \equiv \ln D$ for a range of $q$ upto $\sim q^{*}$. Here $\ln D$ is a proxy for temperature $T$ to reduce the effect of noise in data, and $\tau \equiv \tau_{1/e}$ \Revision{in \textbf{(a)} and $\tau \equiv \tau_{0.25}, \tau_{0.1}$ in \textbf{(b),(c)} respectively.} By definition, $y=1$ at high $T$ but shows a $q$-dependent deviation from 1 at low $T$ revealing the non-Fickian dynamics due to emergent mesoscale correlation. Solid lines are fits of the form $y(x) = 1 + a (x - x_0)^2$ where $a$ is a fit parameter and $x_0$ represent a reference high $T$ ($\ln D$). The Fickian to non-Fickian cross-over temperatures at a given $q$ are extracted by determining where the fit lines intersect $y=1.1$, allowing for a 10\% uncertainty to account for statistical noise in data \cite{ChongKob2009}.
      \textbf{\Revision{(d):}} The cross-over temperatures as a function of $q$, obtained with different probe timescales $\tau_{1/e}, \tau_{0.25}, \tau_{0.1}$ respectively..
      \textbf{\Revision{(e):}} $T$ dependence of the Fickian to non-Fickian cross-over lengthscale obtained by plotting $l_{F} = \frac{2\pi}{q_F}$ as a function of the crossover temperatures. 
    }
    \label{fig:lonset}
\end{figure*}

If the system obeys Fick's law of diffusion \emph{and} the displacement distribution (van Hove function) is Gaussian, then it follows that the self intermediate scattering function $F_s(q,t)$ is Gaussian in $q$ and exponential in time $t$ \cite{HansenMcdonald}, 
\begin{align}
  F_s(q,t) &\equiv \left\langle \frac{1}{N}\sum_{i=1}^N \exp(-\mathrm{i} \vec{q}\cdot [\vec{r}_i(t)-\vec{r}_i(0)])\right\rangle \nonumber\\
  &= \exp(-Dq^2t) \nonumber\\
    D(T)\,q^2\,\tau(q,T) &= \mbox{constant}\;\;\mbox{(Fickian)}
    \label{eqn:Fickian}
\end{align}
Note that in SW silicon we observed no significant difference in the behavior of $\tau(q)$ if $F_s(q,t)$ is replaced by $F(q,t)$ which captures the full many particle dynamics. Combining Eqn. \ref{eqn:Fickian} with Arrhenius laws $D(T) = D_0 \exp(-\frac{E_D}{k_B T})$ and $\tau(q,T) = \tau_0 (q) \exp(\frac{E_\tau(q)}{k_B T})$ we get,
\begin{align}
  D(T)\, q^2\, \tau(q,T) &= D_0\, \tau_0(q)\, q^2 \exp(\frac{\Delta E (q)}{k_B T}) \nonumber\\
  \Delta E (q) &\equiv E_\tau (q) - E_D \nonumber\\
  &= 0, \, \forall q\; \Rightarrow\, \mbox{Fickian.}
  \label{eqn:genFick}
\end{align}
Eqns. \ref{eqn:Fickian}, \ref{eqn:genFick} provide a framework to understand the coupling and decoupling of $D$ with $\tau(q)$. If Eqn. \ref{eqn:Fickian} is obeyed, then $D$ and $\tau(q)$ are mutually coupled \emph{for all $q$}, and the structural relaxation at a lengthscale $q^{-1}$ is due to single particle diffusion with a timescale $\tau_D (q) \propto (D\,q^2)^{-1}$ (Fickian regime). 

The breakdown of Eqn. \ref{eqn:Fickian} indicates a cross-over to a non-Fickian regime. In simple liquids it has been found that due to the emergent dynamical heterogeneity (DH) \cite{Berthier2004, BCG2005, Chong2008, ChongKob2009}, there is a characteristic wave number $q_{F}(T)$ such that for $q < q_{F}(T)$ there is no SEB, but $D$ and $\tau(q)$ are decoupled above $q_{F}(T)$ \cite{TK1995, CE1996, Silescu1999, Ediger2000, Sengupta2013, ChongKob2009, 2014Sengupta, 2016Bhowmik}. The growing many-body correlation due to the DH can be quantified by a characteristic dynamical correlation length \cite{Harrowell2011, KSD2014, KSD2016, Chakrabarty2017}. Among network-forming liquids, the DH is well-documented for water \cite{Becker2006, Mathroo2006, Kumar2007, Mazza2007, Teboul2008,  Banerjee2009, Xu2009, Qvist2012, Rozmanov2012, Dehaoui2015, Henritzi2015, Galamba2017, KK2017, Hijes2018, SRT2018, Ren2018, KK2019,Tsimpanogiannis2020, Dubey2021, Gallo2021} and silica \cite{Kerrache2003, Hoang2006, Nascimento2006, Hoang2007, Saksaengwijit2006, Sen2008, Mallamace2010}, but not so well-reported in silicon. Connection between DH and Fickian to non-Fickian crossover can be directly made by observing that the characteristic lengthscale of Fickian to non-Fickian crossover $l_{F} = \frac{2\pi}{q_F}$ is proportional to the lengthscale of DH independently obtained by other means \cite{Parmar2017}. Note that, existence of such a lengthscale implies that spatial homogeneity or heterogeneity of dynamics at a given temperature should depend on the lengthscale probed, with $D$ and $\tau(q)$ being decoupled at lengthscales below $l_{F}$ \cite{FT2009, FT2011, FT2012}.

We now proceed to analyze the Fickian to non-Fickian crossover in SW silicon along the $P=0$ isobar. In Fig. \ref{fig:RawdataTimeP0}(a)-(c), we show $F(q,t)$ at representative $T$ and $q$. The red and blue data-sets denote temperatures above and below $T_{onset,D}$ respectively. On one hand at small $q \rightarrow 0$, we find Fickian behaviour - exponential decay of $F(q,t)$ - at both high and low temperatures, Fig. \ref{fig:RawdataTimeP0}(a), (here the smallest wave number $q_{min} = \frac{2\pi}{L}$ where $L$ is the simulation box dimension).
On the other hand at large $q \rightarrow q^{*}$, Fig. \ref{fig:RawdataTimeP0}(c), the low $T$, supercooled liquid show two-step, stretched exponential decay, {\it i.e.} non-Fickian dynamics. These are the typical behaviour and similar to supercooled liquids with isotropic interactions. However, in silicon we additionally observe an intermediate regime of $T$ and $q$ where the decay of $F(q,t)$ is single step {\it i.e.} with no clear separation of timescales, but \emph{non}-exponential, see Fig. \ref{fig:RawdataTimeP0}(b). Note that this behaviour is seen not only below $T_{onset,D}$ but also at higher temperatures. These different types of decay are further characterized by the $T$ dependence of the stretched exponent $\beta_{KWW}$ for different $q$ in Fig. \ref{fig:RawdataTimeP0}(d). The presence of the intermediate regime is an indication that the dynamics in SW silicon is somewhat different from that of simple liquids in the Arrhenius regime. 

To ascertain it, we test the validity of Eqn. \ref{eqn:Fickian} in Fig. \ref{fig:RawdataTimeP0}(e) \cite{Chong2008}. Surprisingly, for large $q \rightarrow q^{*}$, the product $Dq^2\tau(q)$ does not become constant even in the Arrhenius regime, which is \emph{a priori} expected to be Fickian. Further, we show $D$ {\it vs.} $\tau(q)$ in Fig. \ref{fig:RawdataTimeP0}(f) main panel. As $q$ increases, we observe two regimes with differing exponent values, but the high $T$ \emph{Arrhenius} regime shows a \emph{decoupling}: $\xi_h \neq 1$ (inset of Fig. \ref{fig:RawdataTimeP0}(f)), demonstrating that the system dynamics can be \emph{non}-Fickian even in the high $T$, \emph{Arrhenius} regime. What might be the reason behind such behaviour at large $q \sim q^{*}$ and high $T$? In silicon, there is a tight correlation between local structure {\it e.g.} local coordination number and mobility \cite{Vasisht2013}. We speculate that the dynamics at the lengthscale $\sim \frac{2 \pi}{q^{*}}$, comparable to the nearest neighbour distances, reflects the local structural heterogeneity, \emph{even} at relatively higher $T$, as our data shows. 

The effect of dynamical heterogeneity should reduce at timescales longer than $\tau_\alpha$ \cite{Parmar2017, 2023Pareek}. Hence, we analyze the $q$-dependence of the dynamics at increasingly longer timescales  $\tau_{1/e} \rightarrow \tau_{0.25} \rightarrow \tau_{0.1}$ which are defined by $F(q,\tau_{p}) = p$, with $p=\frac{1}{e}, 0.25, 0.1$ respectively. The $T$ dependence of these timescales for a range of $q$ are shown in Fig. \ref{fig:qdepEnergy}(a)-(c). According to Eqn. \ref{eqn:genFick}, Fickian behaviour requires that the high $T$ activation energy in the Arrhenius law should be independent of $q$ and equal to $E_D$.  We test this expectation in Fig. \ref{fig:qdepEnergy}(d) for the conventional relaxation timescale, $\tau_{1/e}$, but find that $\Delta E \rightarrow 0$ only in the limit of $q \rightarrow 0$. On the contrary, as $q \rightarrow q^{*}$, $\Delta E$ systematically and significantly deviates from zero. However, as we increase the probe timescale, the $q$ dependence of $\Delta E$ become weaker and almost constant close to zero. This is further verified in Fig. \ref{fig:qdepEnergy}(e) by comparing the power law relationship between $D$ and $\tau_{0.1}$ which shows (inset) that the high $T$ exponent values are now $\approx 1$, in accordance with existing phenomenology. Thus our analysis suggests that the assumption that in the high $T$ \emph{Arrhenius} regime the system dynamics is homogeneous on the time scale of $\tau_{\alpha}$ should be scrutinized carefully in network forming liquids.

\subsubsection{Cross-over lengthscale}
Eqn. \ref{eqn:genFick} suggests that the quantity $y \equiv \frac{D\tau}{D_0 \tau_0 \exp(\frac{\Delta E}{k_B T})}$ should be 1 at high $T$, {\it i.e.} scaling out the $q$ dependence of the high $T$ activation energy should eliminate the effect of heterogeneity in the Arrhenius regime. Any deviation from 1 should reveal the more interesting non-Arrhenius behaviour at low temperatures that is the signature of supercooled liquid dynamics. We test this hypothesis for $P=0$ isobar and different $\tau_p$, following a procedure similar to Refs. \cite{Chong2008, ChongKob2009}. We plot $y$ {\it vs.} $\ln D$ for different $q$ in Figs. \ref{fig:lonset}(a)-(c). Here $\ln D$, which is a slow varying monotone function of $T$, is used as a proxy for $T$ to reduce noise. We see $y=1$ at high $T$ (by definition) and gradually increases in a $q$-dependent way as temperature (diffusivity) decreases. This low $T$ deviation is a manifestation of the non-Fickian nature of the supercooled liquid dynamics. To extract the Fickian to non-Fickian crossover temperature at a given $q$, we fit the data at each $q$ to a function $y(x) = 1 + a (x - x_0)^2$ where $x \equiv \ln D$ and $a$ is a fit parameter. The form of the function is obtained by demanding $y=1$ for $x \geq x_0$ where $x_0$ is reference point at high $T$ ($\ln D$) and also $y'(x_0) = 0$ for smoothness. 
To account for noise in the data, we allow for 10\% uncertainty to determine the temperature where $y$ deviates from 1 \cite{ChongKob2009}. 
The resultant Fickian to non-Fickian crossover temperatures as a function of $q$ are shown in Fig. \ref{fig:lonset}(d). Equivalently, one may extract a crossover lengthscale $l_{F} (T) = \frac{2\pi}{q_F}$ as a function of (crossover) $T$. In Fig. \ref{fig:lonset}(e) we show that $l_{F}$ grows with lowering of $T$.
We note that this behavior is similar to many other supercooled liquids \cite{ChongKob2009, Harrowell2011, KSD2016, Parmar2017}. However, to the best of our knowledge, this issue has not been looked at in silicon or other network-forming liquids. Also, on the timescale of $\tau_{0.1}$, the dynamics becomes Fickian only in the Arrhenius regime, but at low $T$, the $F(q,t)$ still show \emph{non}-exponential decay, see Fig. \ref{fig:RawdataTimeP0}(c). Hence it is not surprising that we find similar values of $l_{F}$ for the entire range of timescales probed. Since DH decreases at longer timescales, a reduction in $l_{F}$ values is expected at even longer timescales where $F(q,t)$ is exponential even at low $T$ \cite{2023Pareek}. Accessing such timescales are however, computationally demanding and beyond the statistical accuracy of the present study.


\section{Summary and conclusion} \label{sec:conclusion}
The present work studies the dynamics of a network-forming liquid {\it viz.} SW silicon over a broad range of pressures (densities) and temperatures and addresses two issues. First, we systematically examine the relationship among different timescales characterizing glassy dynamics and second, we investigate the influence of anomalies, which are a key feature of network-forming liquids, on the dynamics. To answer these questions we use two methods - (a) we determine the loci on P-T plane of two dynamical crossover temperatures: (i) the temperature of the breakdown of the Stokes-Einstein relation, $T_{SEB}$ and (ii) the temperature of the onset of slow dynamics, $T_{onset}$. In the process we compute shear viscosity for (SW) silicon, for the first time to the best of our knowledge. We also systematically show that in SW silicon, $\tau_\alpha$ is a better proxy to $\eta$ rather than $\frac{\eta}{T}$ in the SE relation. (b) To understand the coupling and de-coupling among timescales, we analyze the Fickian to non-Fickian cross-over in dynamics by probing (supercooled) liquid state dynamics at different lengthscales (wave number $q$) and determine the lengthscale of the cross-over. 

The inter-particle interaction in silicon is anisotropic, and it is a poor glass-former. Yet, we find that the dynamics of the supercooled SW silicon is qualitatively similar to good glass-formers with isotropic interactions such as the Kob-Andersen model for colloidal glass-formers. However, in contrast to the colloidal and metallic glass-formers, the loci of $T_{SEB}$ and $T_{onset}$ are clearly different with $T_{SEB} > T_{onset}$. Thus we show that \emph{in silicon, there is a high  $T$ regime where the dynamics is Arrhenius but non-Fickian}. The $q$-dependent analysis suggest that the dynamics is heterogeneous at the nearest neighbour lengthscale - even at high $T$. Presumably this is a manifestation of the local structural heterogeneity due to the anisotropy of the interaction. As the temperature decreases, the onset of slow dynamics occur in the anomalous regime of the P-T plane, approximately around the TMD line indicating the strong influence of anomalies on the supercooled liquid dynamics. \Revision{We show that in this \emph{low} $T$ regime, the dynamics is Fickian only above a certain characteristic lengthscale which grows with temperature.} We show that a correction due to $q$ dependence of the high $T$ activation energy is necessary to accurately estimate this crossover lengthscale. We also provide evidences that this lengthscale is same as that of the dynamical heterogeneity. 

Thus we show that the anisotropy of interaction makes the dynamics of network-forming liquids richer than colloidal and metallic glass-formers. It will be interesting to connect the observed dynamic features to a more microscopic structural picture such as the two-state model, which however, we leave for a future study.  


\begin{acknowledgements}
\Revision{We thank the National Supercomputing Mission (NSM, via grant DST/NSM/R\&D HPC Applications/2021/29)) for providing the computational facilities ``PARAM Yukti'' at Jawaharlal Nehru Centre for Advanced Scientific Research (JNCASR), Bengaluru, India and ``PARAM Ganga'' at the Indian Institute of Technology Roorkee, India, implemented by C-DAC, India and supported by MeitY and DST, Government of India. V.V.V. would like to acknowledge the Bhaskara and the Rahman computational facilities at JNCASR. S.S. thanks the Indian Institute of Technology Roorkee for providing computational support via the Faculty Initiation Grant. H.R. acknowledges junior and senior research fellowships from Council of Scientific Research, India. We would like to express our sincere gratitude to Srikanth Sastry for his invaluable guidance in initiating this work and for his insightful suggestions. We also would like to thank Sarika Maitra Bhattacharyya and Saroj Nandi for their helpful insights.}
\end{acknowledgements}

\section*{Author Declaration}
\subsection*{Conflict of interest}
The authors have no conflicts of interest to disclose.

\subsection*{Author Contribution}
\textbf{Himani Rautela}: Formal analysis (lead); Investigation (lead).
\textbf{Shiladitya Sengupta}: Conceptualization (equal); Supervision (lead); Writing - original draft (equal).
\textbf{Vishwas V. Vasisht}: Conceptualization (equal); Supervision (supporting); Writing - original draft (equal).

\section*{Data Availability}
The data are available from the corresponding author upon reasonable request.

\section*{\Revision{Appendix}}\label{sec:app}

\begin{figure}[htbp]
    \centering
    \includegraphics[keepaspectratio, width=0.30\textwidth]{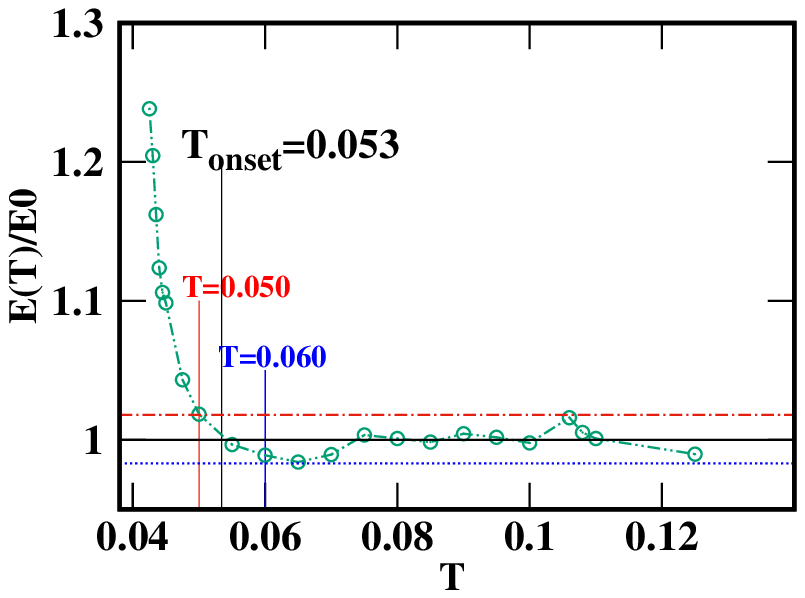}
    \caption{\Revision{\emph{Estimation of uncertainty of $T_{onset}$ values.} We estimate uncertainty in the $T_{onset}$ values using the diffusion coefficient ($D$) along $P=0.0$ isobar as representative data. We define the temperature dependent activation energy $E(T) = T \ln [D0/D(T)]$ and plot the ratio $E(T) / E0$ {\it vs.} $T$. Here $E0$ is the limiting, constant, high-T value in the Arrhenius regime. $T_{onset}$ is the temperature at which the activation energy ratio deviates from 1. The horizontal red and blue horizontal lines are drawn to completely contain the fluctuation in the high temperature data. They are marked at 1.02 and 0.98 respectively. Thus allowing for 2\% uncertainty in the input diffusion coefficient data, we estimate upper and lower bounds of Tonset where $E(T)/E0$ crosses red and blue lines respectively. This yields a range of Tonset ~ [0.50 - 060] along P=0.0 isobar.}}
    \label{fig:Tonset_errorbar}
\end{figure}

\begin{figure*}[htbp]
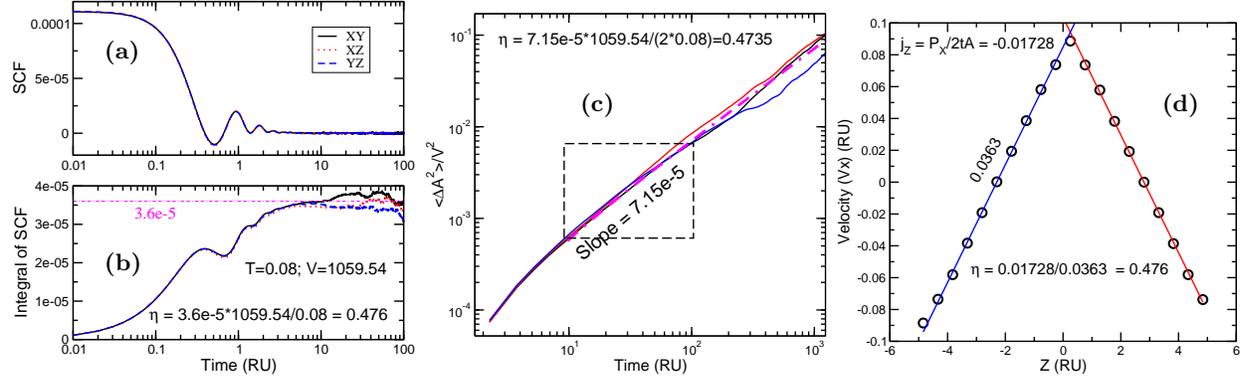

    \centering
    \includegraphics[keepaspectratio, width=0.30\textwidth]{SCF-Eta-T0.08-P0.0.eps}
    \includegraphics[keepaspectratio, width=0.30\textwidth]{Einstein-Eta-T0.08-P0.0.eps}
    \includegraphics[keepaspectratio, width=0.30\textwidth]{VelProf-MP-Eta-T0.08-P0.0.eps}
    \put(-430,120){\textbf{(a)}}
    \put(-430,40){\textbf{(b)}}
    \put(-250,100){\textbf{(c)}}
    \put(-30,100){\textbf{(d)}}
    \caption{\Revision{\emph{Details of shear viscosity calculations at a representative state point.} \textbf{(a):} Time dependence of the stress correlation function (SCF) for Green-Kubo (GK) method. SCF computed along different orientations overlap with each other, as expected in equiibrium. \textbf{(b):} Integral of stress correlation function converges to a plateau in the long time limit. \textbf{(c)} Time dependence of the variance of the Helfand moment for Einstein method, highlighting linear regime. \textbf{(d):} Testing linearity of the imposed velocity profile for the NEMD method. Data shown for $P=0.0$ and $T=0.8$ (both in R.U.). Details of $\eta$ calculations by each method are shown inside panels. }}
    \label{fig:eta_det}
\end{figure*}

\clearpage
\bibliographystyle{plain}

\end{document}